\documentclass[10pt,aps,prx,twocolumn,tightenlines,amsmath,amssymb,superscriptaddress]{revtex4-2}
\usepackage{textcomp,pbox}
\usepackage[export]{adjustbox}
\usepackage{bm}
\usepackage{dcolumn,booktabs,url}
\usepackage[scaled]{helvet}
\usepackage{orcidlink}
\usepackage{sansmath,gensymb}
\usepackage{graphicx,color}
\usepackage{multirow}
\usepackage{physics}
\usepackage{pdfpages}
\usepackage{array}
\usepackage{makecell}
\usepackage{tabularx}
\usepackage{dsfont}
\usepackage{hyperref}
\hypersetup{
     colorlinks   = true,
     citecolor    = red,
     linkcolor    = blue,
     urlcolor     = red     
}

\newcommand{\fref}[1]{\text{Fig.}~\ref{#1}}

\makeatletter
\AtBeginDocument{\let\LS@rot\@undefined}
\makeatother

\begin{document}
\title{Directional and correlated optical emission from a waveguide-engineered molecule with local control}

\newcommand{\AffCPH}{Center for Hybrid Quantum Networks (Hy-Q), Niels~Bohr~Institute, University~of~Copenhagen, Jagtvej 155A, DK-2200 Copenhagen, Denmark}
\newcommand{\AffBochum}{Experimental Physics VI, Faculty of Physics and Astronomy, Ruhr University Bochum, Universit\"atsstra\ss e 150, 44801 Bochum, Germany}
\newcommand{\AffVien}{Vienna Center for Quantum Science and Technology (VCQ) and Christian Doppler Laboratory for Photonic Quantum Computer, University of Vienna, 1090 Vienna, Austria}

\author{Clara Henke\,\orcidlink{0009-0004-5255-5689}}
\thanks{These authors contributed equally to this work.}
\author{Thomas Wilkens Sandø\,\orcidlink{0009-0008-5931-2910}}
\thanks{These authors contributed equally to this work.}
\author{Vasiliki Angelopoulou\,\orcidlink{0000-0002-8115-3169}}
\affiliation{\AffCPH{}}
\author{Lena Maria Hansen\,\orcidlink{0009-0003-9352-6246}}
\affiliation{\AffVien{}}
\author{Alexey Tiranov\,\orcidlink{0000-0003-0791-8730}}
\author{Oliver August Dall'Alba Sandberg\,\orcidlink{0000-0002-9410-6150}}
\author{Zhe Liu\,\orcidlink{0000-0003-0672-4328}}
\author{Leonardo Midolo\,\orcidlink{0000-0003-0237-587X}}
\affiliation{\AffCPH{}}
\author{Nikolai Bart}
\author{Arne Ludwig\,\orcidlink{0000-0002-2871-7789}}
\affiliation{\AffBochum{}}
\author{Anders Søndberg Sørensen\,\orcidlink{0000-0003-1337-9163}}
\author{Peter Lodahl\,\orcidlink{0000-0002-9348-9591}}
\author{Cornelis Jacobus van Diepen\,\orcidlink{0000-0001-8454-2859}}
\email[Email to: ]{cjvandiepen@gmail.com}
\affiliation{\AffCPH{}}

\begin{abstract}
Radiative coupling between quantum emitters leads to a range of spectacular emission phenomena. Dicke studied the foundations of collectively enhanced and suppressed decay, commonly referred to as super- and subradiance. Collective effects can further result in directionality of the emission, thus offering a complimentary implementation of chiral quantum optics. Waveguide quantum electrodynamics (QED) allows coupling between spatially separated emitters, enabling selective driving. In this work, we control the emission direction for a pair of quantum dots embedded in a bidirectional photonic crystal waveguide offering independent electrical tuning. Notably the emitters are 13 \micro m apart, which corresponds to 26 effective wavelengths, but are nevertheless radiatively coupled. The directionality arises from a dispersive dipole-dipole interaction, which shifts the energy of the collective states, so that the emitter pair effectively forms an artificial molecule. We show that the emission direction can be switched from left- to rightwards by manipulating the relative driving phase while collectively exciting the emitters. In addition, we observe directional photon statistics under continuous driving, with, for example, single photons detected on one output port, and photon pairs on the other. With pulsed excitation, both emitters are fully inverted and correlated photon pairs are observed in time-resolved intensity correlation measurements. This work demonstrates a novel implementation of chiral quantum optics using quantum dots coupled via a non-chiral waveguide, and reports key steps for scaling up as a multi-emitter waveguide QED platform. 
\end{abstract}

\maketitle

\section{Introduction}
Quantum networks~\cite{Kimble2008,Wehner2018} form the backbone for distributed quantum computing~\cite{Daiss2021}, entangled atomic clocks~\cite{Nichol2022}, and quantum key distribution~\cite{Xu2020}. Their capabilities rely on the transfer of quantum information between distant nodes. The connection between nodes is typically achieved via room temperature fibers, which necessitates encoding quantum information into optical photons to suppress thermal noise. The demand for efficient routing of these photons requires control over their emission from quantum light sources. This emission control can be achieved by tailoring the environment~\cite{Purcell1946} with photonic nanostructures. Waveguide QED provides a natural framework for the generation and manipulation of propagating photons.~\cite{Ciccarello2024, Sheremet2023}. 

In typical waveguide structures, photons are emitted to each output port with equal probability, but this symmetry can be broken by carefully engineered structures. Propagation-direction dependent light-matter interaction is the basis of chiral quantum optics~\cite{Pichler2015,Lodahl2017}. Examples include the polarization-dependent scattering with a gold nanoparticle on a nanofiber~\cite{Petersen2014}, directional emission from cesium atoms laser-trapped near a waveguide~\cite{Mitsch2014}, a chiral giant atom~\cite{Joshi2023} and a quantum dot in a reflection-symmetry broken photonic crystal waveguide~\cite{Sollner2015}. Alternatively, multiple waveguide-coupled emitters that effectively form an artificial molecule have been proposed for directional emission~\cite{Gheeraert2020,Guimond2020}, and this has been realized in the microwave regime~\cite{Kannan2020,Aamir2022,Redchenko2023}. The required level of control over multiple coupled emitters has so far, however, not been realized in the optical domain. 

Emission interference via coupling to a common radiative mode can lead to super- and subradiance~\cite{Dicke1954,Gross1982}. The coupling of distant emitters is enabled by the spatially extended mode realized in waveguide QED~\cite{Sheremet2023}. Quantum dots in waveguides~\cite{Lodahl2015} are of particular interest because of their strong coupling to the guided mode~\cite{Arcari2014}. Over recent years, several works have reported multiple resonant quantum dots in a single waveguide~\cite{Kim2018, Grim2019, Chu2023,Hallacy2025}. Furthermore, super- and subradiance have been observed in the time-resolved emission from waveguide-coupled quantum dots~\cite{Tiranov2023}. Even when the waveguide is bidirectional, the collective emission from multiple emitters can be strongly directional due to interference effects. For optical emitters, directionality in scattering has been observed with a pair of silicon vacancies in a microdisk resonator~\cite{Lukin2023}, but lacked the capacity for selective driving to control the direction of emission. 

In this work, we demonstrate and control directionality of emission from a pair of waveguide-coupled quantum dots. First, the emitters are brought into resonance by independently tuning their dc Stark shifts using local electric fields. Next, the time-resolved emission dynamics show waveguide-mediated coupling between the emitters, and that this coupling has a dispersive component. Then, by manipulating the relative phase between the driving fields, we control the direction of single-photon emission in the pulsed collective-excitation regime. Additionally, the photon statistics is studied under continuous driving, which show predominantly single photons at the left end of the waveguide and photon pairs at the right end. Finally, we simultaneously perform Rabi $\pi$-flips to fully excite both emitters and observe correlations and directionality in the emission of photon pairs. 
Together, these results show that waveguide-coupled quantum emitters constitute a versatile platform for routing optical photons and generating correlated states of light. 

\section{Platform for multi-emitter directional emission}
\begin{figure*}
	\includegraphics[width=\textwidth]{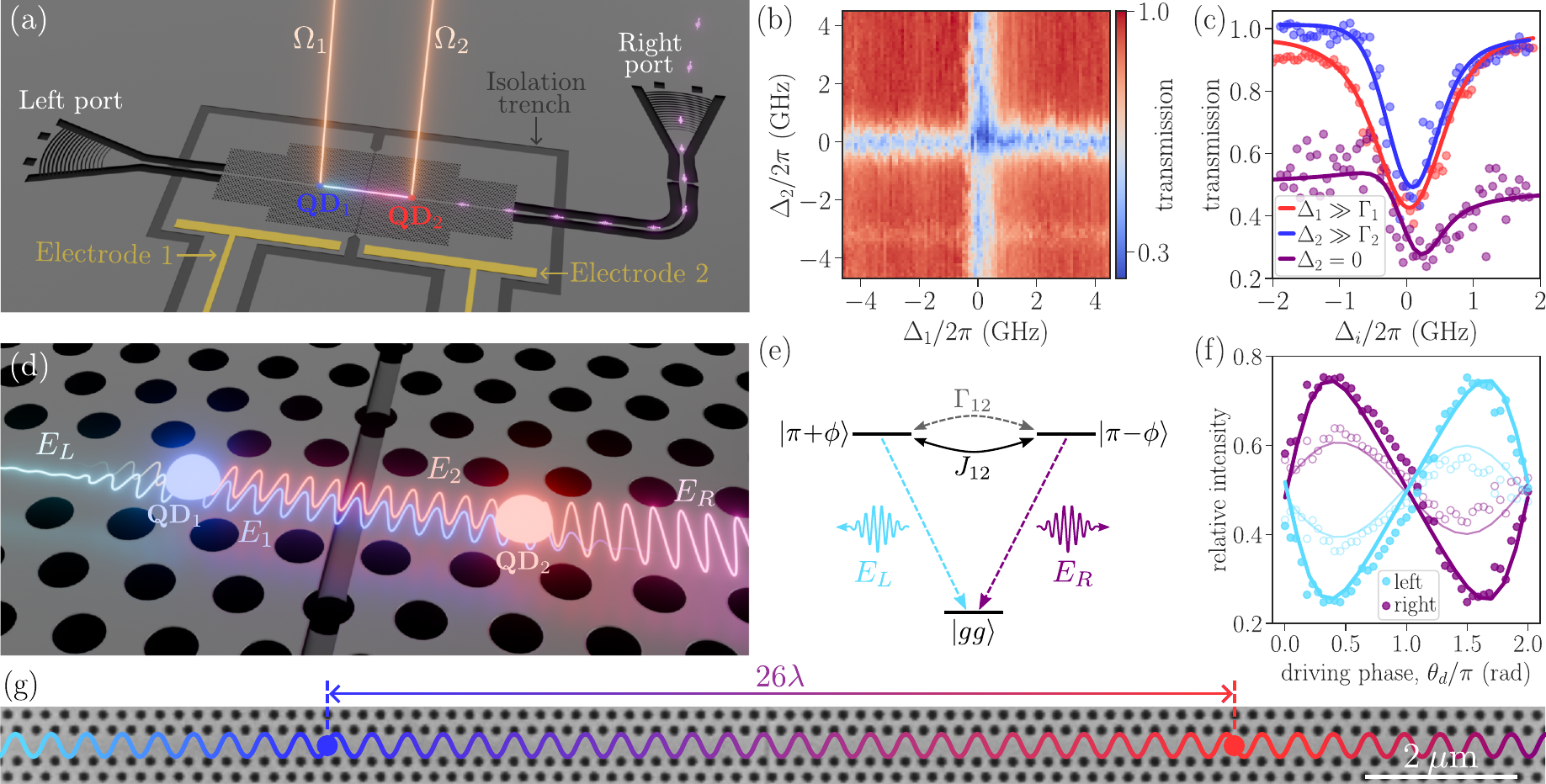}
	\caption{ \textbf{Coupled distant emitters with independent tuning and directional emission.} 
    (a) Rendered image of the photonic crystal waveguide (PCW). QD$_1$ and QD$_2$ can independently be tuned electrically based on the isolation trench and the separate electrodes. Optical driving fields are labeled $\Omega_1$ and $\Omega_2$. For clarity, the right port is visualized closer than in the actual device. (b) Waveguide transmission as a function of the detuning of both emitters. The detunings are controlled through the bias voltages that dc Stark shift the emitters. The faint additional horizontal line corresponds to the other dipole of QD$_2$. (c) Waveguide transmission as a function of detuning. The blue and red datapoints are detuning-dependent single emitter data (either QD$_1$ or QD$_2$) where the other QD is detuned far away from the laser frequency. The purple datapoints are taken with QD$_2$ on resonance with the laser, and detuning QD$_1$. Lines are fits to the data. (d) Zoom-in of (a) on the PCW region that hosts two QD emitters shown as half spheres. The QDs are coupled via the waveguide mode and directional emission is a result of constructive/destructive interference. The distance between the QDs compared to the photonic crystal lattice is in reality ten times larger than what is depicted. (e) Energy level diagram for two waveguide-coupled QDs with decay from the single-excitation subspace. Photons are emitted either to the left or to the right by decay of the states $\ket{\pi\pm\phi} = \frac{1}{\sqrt{2}} \left( \ket{eg} - e^{\pm i\phi} \ket{ge} \right)$. (f) Directionality in emission characterized by the relative intensities, $I_{L/R}/(I_L+I_R)$, as a function of the relative driving phase. The emitters are collectively driven with equal Rabi frequencies, $|\Omega_1| = |\Omega_2|$, using pulsed resonant excitation. Filled (open) circles correspond to the total emission integrated over $0.4$ ns ($3$ ns) after the excitation pulse and solid lines show corresponding numerical simulation results. (g) Scanning electron microscopy image zoomed in on the waveguide in the photonic crystal with in the middle the trench faintly visible. Representative QD positions, obtained from imaging of the device (see Appendix~\ref{app:sample_setup}), are indicated by colored disks. The field is colored according to the QD positions and emission directions.
    }
	\label{fig:1}
\end{figure*}

The device, visualized in Fig.~\ref{fig:1}(a), contains a bidirectional photonic crystal waveguide patterned in a suspended membrane. The membrane is formed out of a heterostructure based on GaAs and acts as a p-i-n diode. Light can be coupled into and out of the device via grating coupling ports at each end of the waveguide. Embedded in the waveguide are InGaAs quantum dots (QDs), which can act as single-photon emitters. Their emission frequency is electrically tunable through the dc Stark effect using the p-i-n diode. A shallow trench etched through the p-doped layer enables independent tuning of emitters in the left and right halves of the waveguide using locally controlled electric fields~\cite{Papon2023}. More details on the device and heterostructure are provided in Appendix~\ref{app:sample_setup}.

We operate a pair of QDs, one in each half of the waveguide, whose neutral exciton transitions can be tuned into resonance. Each QD is treated as a two-level system with a ground state, $\ket{g}$, and an excited state, $\ket{e}$. The QDs can be individually addressed optically from free space, facilitated by their 13 \micro m separation. This corresponds to a distance of 26 wavelengths, see \fref{fig:1}(g), as compared to the wavelength at the band edge of the PCW where $\lambda = 2a = 500$ nm, with $a$ the photonic crystal lattice constant. The coupling of the emitters to the waveguide mode is characterized by $\beta_m=\gamma^{wg}_m/\Gamma_m$, where $\gamma^{wg}_m$ is the decay rate into the waveguide and $\Gamma_m$ is the total decay rate for emitter $m$. The total decay rates, measured individually while detuning the other emitter far off resonance, are $\Gamma_1/2\pi=0.388(1)$ GHz and $\Gamma_2/2\pi=0.349(1)$ GHz (see Appendix~\ref{app:qd_char}). Collective excitation of the QDs is performed with a spatial light modulator (SLM). Using the SLM, a single laser beam is diffracted into separate beams with control over the relative beam spot positions, driving strength and phase (see Appendix~\ref{app:slm}).

Despite the large distance between the emitters, their coupling to a common guided mode results in an effective interaction between them. This interaction can have dissipative, $\Gamma_{12} = \sqrt{\gamma^{wg}_1 \gamma^{wg}_2} \cos{\phi}$, and dispersive, $J_{12} = \frac{1}{2}\sqrt{\gamma^{wg}_1 \gamma^{wg}_2} \sin{\phi}$, coupling components, dependent on the emitter-emitter coupling phase, $\phi=kx$, where $x$ is the distance between the emitters and $k$ is the effective wave number for the guided mode. The dispersive coupling results in a splitting of the energy levels for the collective states $\ket{\pm}=\left(\ket{eg} \pm \ket{ge} \right)/\sqrt{2}$, and their decay rates are modulated by the dissipative component, which results in enhanced and suppressed emission. 

The coupling of the emitters to the guided mode is experimentally studied by sending a weak continuous-wave laser field into the waveguide via the left port and collecting the light from the right port, while electrically varying the emission frequencies of the QDs. Figure~\ref{fig:1}(b) shows the measured fraction of transmitted intensity as a function of the detuning, $\Delta_i$, of each of the emitters, i.e., the difference in the QD emission frequency with respect to the laser field frequency, which here is 320.861~THz. We observe a reduction in the transmission when tuning in resonance with either of the QDs leading to a ``cross pattern". Each line corresponds to the individual resonances of QD$_1$ and QD$_2$, which act as mirrors selectively reflecting a single photon at a time. At the intersection, where both emitters are in resonance with the laser, the transmission is lower than for the single emitters, i.e., multiple quantum emitters can reflect more efficiently than a single emitter~\cite{Chang2012}. The absence of crosstalk in \fref{fig:1}(b) confirms the independent electrical tuning of the QDs. 

Figure~\ref{fig:1}(c) shows linecuts of \fref{fig:1}(b) with the fraction of transmitted intensity as a function of detuning for one of the emitters. Initially, either emitter is tuned into resonance with the laser field while the other emitter is far detuned, essentially characterizing the emitters individually. Dips in the transmitted intensity are found with a relative depth of $I^\mathrm{dip}_1/I_\mathrm{in} = 0.492(4)$ for QD$_1$ and $I^\mathrm{dip}_2/I_\mathrm{in} = 0.426(4)$ for QD$_2$, where $I_\mathrm{in}$ is the intensity of the laser light coupled into the waveguide. This shows that both emitters are well coupled to the guided mode, i.e., that their $\beta$-factors are high (close to unity)~\cite{Arcari2014}. Then, the second emitter is fixed on resonance with the laser, and the first emitter is tuned into resonance as well. This results in a transmission dip with depth $I^\mathrm{dip}_{12}/I_\mathrm{in}=0.239(3)$, which is deeper than for the individual emitters. It is shallower than the product of the two emitter dips, $I^\mathrm{dip}_1I^\mathrm{dip}_2/I^2_\mathrm{in}=0.210(3)$, which indicates that the emitters are coupled. The transmission dip with both emitters together is found at a small detuning, and is asymmetric in the detunings of the individual emitters. These constitute a signature of a dispersive component in the coupling between the emitters, which shifts the energy of their collective states, hence they form a molecule. 

The dispersive coupling is inherently linked to interference that results in directionality of the emission, as visualized in \fref{fig:1}(d). The emission direction depends on two phases in the system, the coupling phase, $\phi$, determined by the positions of the emitters, and the relative phase, $\theta_d$, between the driving fields of the two emitters. The fields emitted to the left, $E_L$, and to the right, $E_R$, into the waveguide (assuming $\gamma^{wg}_1=\gamma^{wg}_2=\gamma^{wg}$) are 
\begin{align}
E_L & = i\sqrt{\frac{\gamma^{wg}}{2}}\left( \sigma^-_1 + e^{i\phi} \sigma^-_2 \right), \\
E_R & = i\sqrt{\frac{\gamma^{wg}}{2}}\left( \sigma^-_1 + e^{-i\phi} \sigma^-_2 \right), 
\end{align}
where $\sigma^-_m$ is the Pauli lowering operator for emitter $m$. For the state $(\ket{eg} + e^{i\theta_d}\ket{ge})/\sqrt{2}$, the emitted fields are
\begin{alignat}{2}
    E_L & \propto e^{i0} e^{i0} + e^{i\theta_d} e^{i\phi} && = 1 + e^{i(\theta_d +\phi)}, \\
    E_R & \propto e^{i0} e^{i0} + e^{i\theta_d} e^{-i\phi} && = 1 + e^{i(\theta_d -\phi)}.
\end{alignat}
If $\theta_d = \pi - \phi$, interference cancels the left-going field, $E_L = 0$. Conversely, for $\theta_d = \pi + \phi$, the right-going field vanishes, $E_R = 0$. Both of these are depicted by the energy level diagram in \fref{fig:1}(e). Given that $\phi \neq N \pi$, the emission in either one of the directions can thus be suppressed by controlling the driving phase, hence switching the emission direction. 

To realize on-demand directional emission, both emitters are driven from free space with pulsed resonant excitation. We make use of an SLM to create two beams with mutual coherence and relative phase control. The driving fields are set to be relatively weak such that emission is predominantly from the single-excitation subspace. Figure~\ref{fig:1}(f) shows the relative intensities, $I_{L/R}/(I_L+I_R)$, measured on the left and the right port compared to the total intensity (see Appendix~\ref{app:muliplex-collection} for details on the setup and Appendix~\ref{app:dir_em} for the time-resolved emission data). By changing the relative phase between the two excitation beams, the emission is switched from being bidirectional to being predominantly to the right or to the left. This explicitly demonstrates control over the emission direction by using the driving phase as a control knob. 

\section{Dispersive coupling in emission dynamics}
\begin{figure}
    \includegraphics[width=\linewidth]{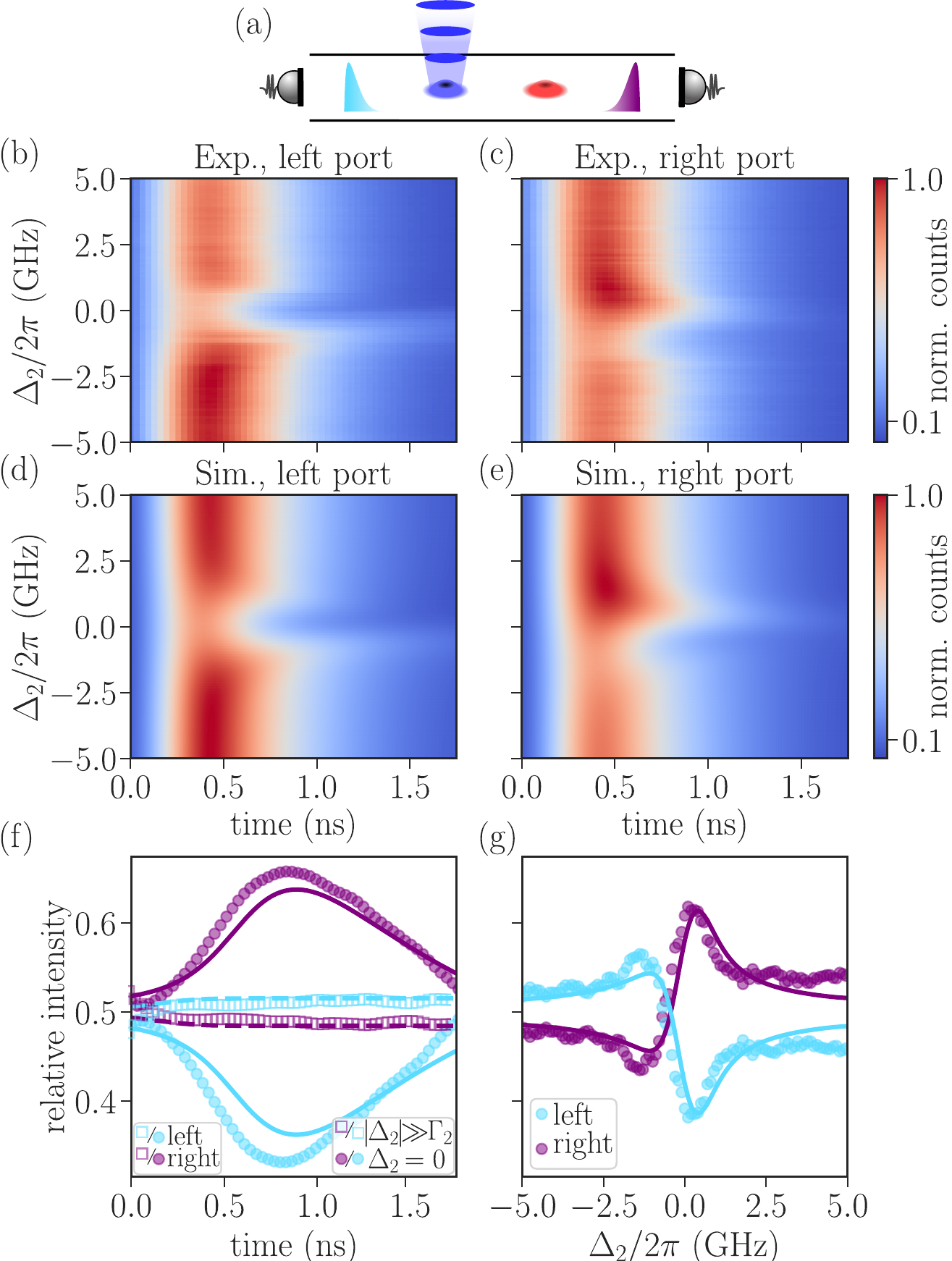}
	\caption{ \textbf{Emission dynamics with dispersive coupling.} 
    (a) Schematic for pulsed excitation of the left emitter and collection on the left and right port. (b), (c) Measured time-resolved emission intensity as a function of the detuning of QD$_2$ for the (b) left and the (c) right port after pulsed resonant excitation of QD$_1$. The counts measured on the left and right port are normalized by the maximum in the full 2D map for the respective side. (d), (e) Numerical simulations corresponding to the experimental results shown in (b) and (c). (f) The relative time-resolved intensity measured on the left (cyan) and right (purple) ports with QD$_2$ resonant (filled circles) or far detuned (open squares). Solid (dashed) lines show numerical simulations with the right emitter resonant (far detuned). (g) Directionality in time-integrated emission ($2$ ns window) as a function of the detuning of QD$_2$. Simulations are shown by solid lines. 
    }
	\label{fig:2}
\end{figure}

Lifetime measurements provide a powerful method to characterize coupled quantum emitters. For multiple emitters, it enables the discrimination between radiatively coupled and independent emitters~\cite{Koong2022}. In particular, the presence of super- and subradiance is directly revealed by increased and decreased decay rates. In this section, we study the coupling character between the QDs using pulsed resonant excitation of QD$_1$ without driving QD$_2$. The excitation power is calibrated to a $\pi$-flip based on a Rabi oscillation measurement, thus initializing the system in the state $\ket{eg}$. The time-resolved emission intensities at the left and right ports of the waveguide are measured with single-photon detectors, as schematically depicted in \fref{fig:2}(a).

Figure~\ref{fig:2}(b) and (c) show the emission intensity collected at the left and right port as a function of the detuning of the undriven emitter. For both ports, a modulation is observed as the undriven emitter is tuned into resonance, which demonstrates that the emitters are radiatively coupled. In addition, there is an asymmetry between left and right ports, and between positive and negative detuning. This contrasts with previous work~\cite{Tiranov2023} where the coupling was of predominantly dissipative character ($\phi=N\pi$), which leads to symmetric behavior.

To assess the coupling character, it is helpful to express the emission in terms of populations in the excited states. The left and right emission intensities (assuming $\gamma^{wg}_1=\gamma^{wg}_2=\gamma^{wg}$ and including only the single-excitation subspace) are $I_L = \langle E^\dagger_L E_L \rangle = \gamma^{wg} p_{-\phi}$ and $I_R = \langle E^\dagger_R E_R \rangle =\gamma^{wg} p_{+\phi}$, with $p_{\pm\phi} = \langle \pm\phi | \rho | {\pm}\phi \rangle$ the population in the state $\ket{\pm\phi}=(\ket{eg} + e^{\pm i\phi}\ket{ge})/\sqrt{2}$. Note that the state $\ket{+\phi}$ ($\ket{-\phi}$) is orthogonal to $\ket{\pi + \phi}$ ($\ket{\pi - \phi}$) and that decay from the latter results in emission only to the left (right). Furthermore, the states $\ket{+\phi}$ and $\ket{-\phi}$ are orthogonal if and only if $\ket{-\phi}=\ket{\pi+\phi}$, i.e., $\phi=(N + \frac{1}{2})\pi$ with $N \in \mathbb{Z}$. Only in this case is their emission fully directional. The experimentally observed asymmetry between emission at the left and right ports directly shows that $p_{+\phi} \neq p_{-\phi}$, and thus $\phi \neq N \pi$. The coupling thus has a dispersive component, $J_{12} \propto \sin \phi$, thereby placing the system in a novel regime of collective emission. 

Numerical simulations for the time-resolved emission at the left and right port are shown in \fref{fig:2}(d) and (e). These results, which include spectral diffusion and detection jitter, show good agreement with the experimental data. The simulations were performed for $\phi=0.8\pi$, thus confirming that the emitters are in a coupling regime with a sizable dispersive component. 

The dynamics of the emission direction with both emitters resonant and with the undriven emitter far detuned, are shown in~\fref{fig:2}(f). For both cases, shortly after the excitation pulse there is no directionality, $I_L=I_R$, because the system is initialized in $\ket{eg}$. With the second emitter far detuned, the emission to the left and the right remains equal, because the emission from the single QD is symmetric. With the second emitter on resonance, the relative emission to the right first increases but eventually becomes symmetric again. This is a consequence of the dispersive coupling, resulting in the initial state evolving towards $\ket{\pi-\phi}$, thus $p_{+\phi} > p_{-\phi}$ and $I_R > I_L$. Simultaneously, as the population in the state $\ket{+}$ decays superradiantly, due to the dissipative coupling, the system evolves towards the slowly-decaying state $\ket{-}$, from which emission is symmetric between both directions again. 

Figure~\ref{fig:2}(g) shows the time-integrated relative emission as a function of detuning for the second emitter. The dominant emission direction switches from left to right as the detuning is changed from negative to positive. This shows that the detuning can be used to switch the emission direction. In future work, leveraging the control over both the detuning and driving phase may enable further control over the emission direction. 

In previous work on radiatively coupled QDs in a PCW~\cite{Tiranov2023}, the coupling was found to be of predominantly dissipative character, i.e., $\phi = N \pi$ with $N \in \mathbb{Z}$. Here, we observe a coupling regime which has both substantial dissipative and dispersive components. We hypothesize that this is enabled by selecting emitters that are at a moderate spectral distance ($3.5$ THz) from the band edge of the PCW (at 317.4 THz) combined with a larger spatial separation between the emitters. At the band edge we have $ka=\pi$, such that near the band edge $ka = \pi - \varepsilon$, with $\varepsilon$ a small number, $k$ the wave number and $a$ the lattice constant. QDs with a high $\beta$-factor have a well defined position within the photonic crystal unit cell~\cite{Javadi2018}, thus the distance, $x$, between a pair of emitters separated by $M$ unit cells can be approximated as $x = M a$. Taken together, the phase lag that accumulates between the emitters is $\phi = k x = M (\pi -\varepsilon)$. The combination of working further from the band edge, leading to a larger value of $\varepsilon$, and the large spatial separation $M\approx 50$ unit cells, leads to a larger coupling phase, which for this emitter pair is found to be $\phi = (0.8 + N) \pi$. Further insight into the coupling between emitters via a photonic crystal waveguide may be obtained from numerical simulations~\cite{Chu2022}.

\section{Directionality in photon statistics}
Next, the photon statistics of the collectively coupled emitters is studied through second-order correlation measurements to record $g^{(2)}$. It signifies, for example, the emission of single photons through the observation of antibunching ($g^{(2)}(0) < 1$)~\cite{Kimble1977}. For a pair of emitters, continuously driven below saturation, correlations in the form of an antidip, i.e., a dip superimposed with a peak, can in principle arise both from radiatively coupled~\cite{Hallett2024} and from uncoupled emitters~\cite{Koong2022,Cygorek2023}. Recent work demonstrated that an antidip in $g^{(2)}(\tau)$ is a distinctive signature of coupling between the emitters in the case that only one emitter is driven~\cite{VanDiepen2025}. However, the role of directionality in the intensity correlations has not been addressed thus far. 
\begin{figure}[h!]
    \includegraphics[width=1\linewidth]{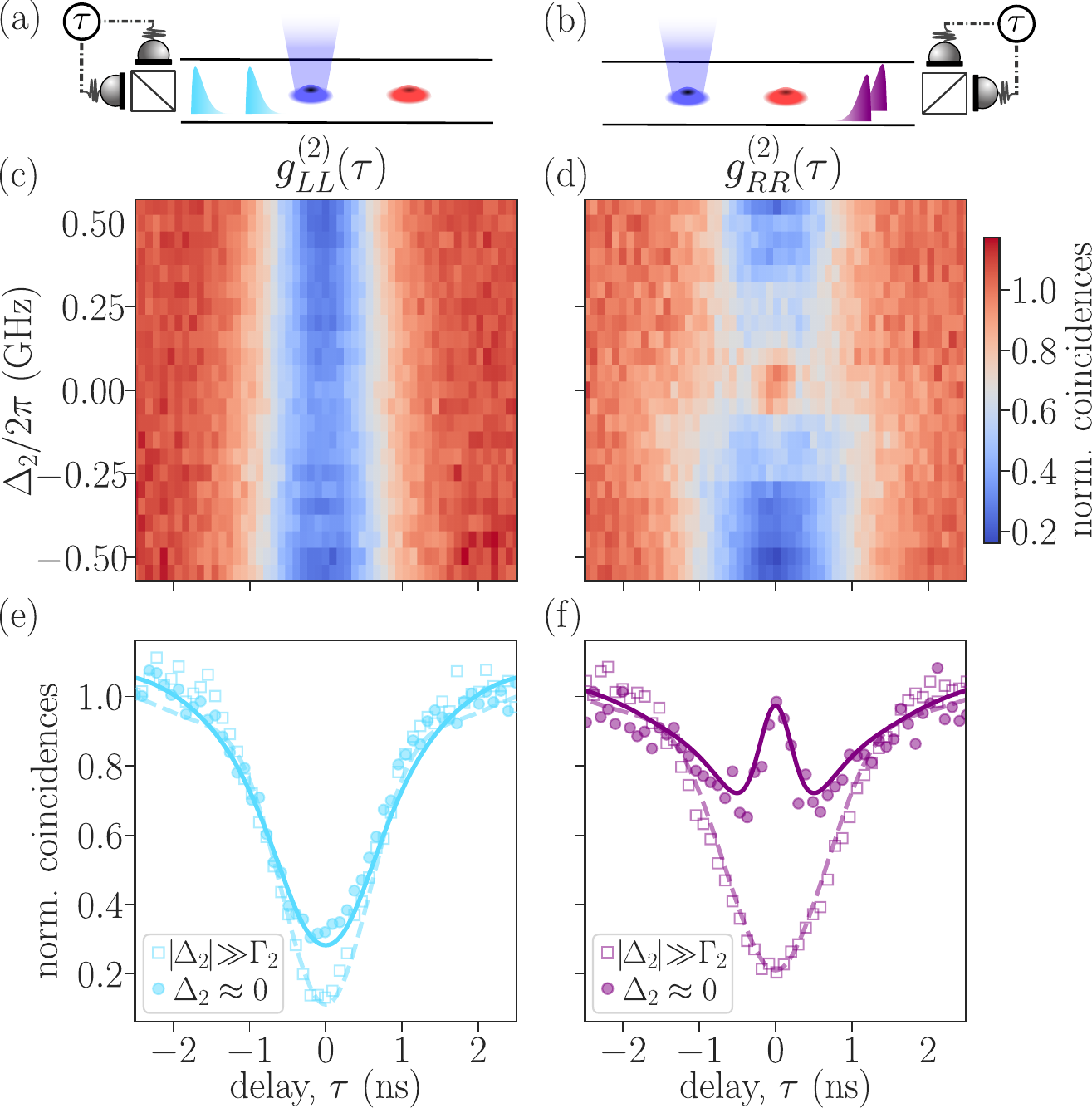}
	\caption{\textbf{Directionality in photon statistics.}
    (a), (b) Schematic of a pair of emitters in a waveguide with the left one continuously driven from free space. At the (a) left or the (b) right port a Hanbury Brown-Twiss setup equally splits the scattered light into two detectors. (c), (d) Intensity correlations measured at the (c) left and (d) right port as a function of the detuning of the undriven QD$_2$ with a continuous and weak resonant drive of QD$_1$. (e), (f) Linecuts from (c) and (d), with the undriven emitter resonant (filled circles) or far detuned (open squares). Solid (dashed) lines show numerical simulations with QD$_2$ resonant (far detuned). 
    }
	\label{fig:3}
\end{figure}

The normalized second-order intensity correlations in terms of the field operators are
\begin{equation}
g^{(2)}_{\alpha\beta}(t,t+\tau) = \frac{\langle E_\alpha^\dagger(t) E_\beta^\dagger(t+\tau) E_\beta(t+\tau) E_\alpha(t) \rangle}{\langle E_\alpha^\dagger(t) E_\alpha(t) \rangle \langle E_\beta^\dagger(t+\tau) E_\beta(t+\tau) \rangle},
\end{equation}
with $\alpha, \beta \in \{ L, R\}$ indicating whether the photon is emitted to the left or right. Under continuous driving, the system reaches a steady state for which the correlations are $g^{(2)}_{\alpha\beta}(\tau) = \lim\limits_{t\rightarrow\infty} g^{(2)}_{\alpha\beta}(t,t+\tau)$. The intensity correlations are measured at the left and right port using Hanbury Brown-Twiss (HBT) setups as schematically depicted in \fref{fig:3}(a) and (b). 

We drive the left emitter resonantly with a weak continuous laser field, while the right emitter is not driven. Figure~\ref{fig:3}(c) and (d) show the intensity correlations measured at the left and the right port as a function of the detuning of the non-driven emitter. Linecuts with both the emitters resonant and the undriven emitter far detuned are shown in \fref{fig:3}(e) and (f) for the left and right port, respectively. While the undriven emitter is far detuned, the intensity correlations on both sides have a characteristic single-emitter anti-bunching dip ($g^{(2)}_{LL}(0) = 0.13, g^{(2)}_{RR}(0) = 0.2$). As the undriven emitter is brought into resonance, a striking contrast arises in the correlations at the two ports. For the left port, the single-emitter dip becomes shallower ($g^{(2)}_{LL}(0)=0.31$), while for the right port, a clear antidip appears ($g^{(2)}_{RR}(0)=0.98$). Note that detection jitter lowers the height of the antidip and therefore reduces the underlying correlation peak. In addition, a temporal broadening of the dip is observed, in particular noticeable in \fref{fig:3}(d), which is attributed to the single photons mainly being emitted from the subradiant state~\cite{VanDiepen2025}. In Appendix~\ref{app:g2_cw}, intensity correlation measurements are shown while collectively driving both emitters in resonance. Dependent on the relative driving phase, either an antidip with a bunching peak ($g^{(2)}_{LL}(0)=1.05$, $g^{(2)}_{RR}(0)=1.71$) or antibunching ($g^{(2)}_{LL}(0)=0.40$, $g^{(2)}_{RR}(0)=0.39$) is measured. 

The directionality in photon statistics can be understood from the relation between intensity correlations and populations
\begin{align}
g^{(2)}_{LL}(0) & = \frac{p_{ee}}{\left(p_{ee} +p_{-\phi} \right)^2} , \label{eqn:g2ll_pops} \\
g^{(2)}_{RR}(0) & = \frac{p_{ee}}{\left(p_{ee} +p_{+\phi} \right)^2} , \label{eqn:g2rr_pops}
\end{align}
with $p_{ee}$ the population in the double-excited state. For a weak continuous and resonant drive of only the left emitter and assuming for simplicity equal coupling to the waveguide mode ($\beta_1=\beta_2=\beta$), the single-excitation component of the steady state is $\ket{\psi_1} \propto \ket{eg} - \beta e^{i\phi} \ket{ge}$ (see Appendix~\ref{app:g2_cw}). In the limit of $\beta \rightarrow 1$, this component becomes $\ket{\psi_1} \propto \ket{eg} - e^{i\phi} \ket{ge} = \ket{\pi + \phi}$. Thus, $p_{+\phi} \rightarrow 0$ and $g^{(2)}_{RR}(0)\rightarrow1/p_{ee}$ which leads to strong bunching on the right for a weak drive. Since $p_{-\phi}$ remains finite, single photons are emitted to the left while photon pairs are emitted equally in both directions leading to the observed asymmetry. This is consistent with QD$_2$ acting as a saturable mirror for the photon source QD$_1$. For the intensity correlations, it follows that $g^{(2)}_{LL}(0) \leq g^{(2)}_{RR}(0)$, which is consistent with our experiment. Simulation results for different coupling phases are provided in Appendix~\ref{app:g2_cw}. 

\section{Full inversion and correlated emission}
Single QDs have been used to generate correlated photons through scattering with a weak coherent pulse~\cite{Jeannic2022,Tomm2024, Liu2023}, and by repeated emission from a charged QD using the spin degree of freedom~\cite{Schwartz2016,Meng2024}. An exciting natural advantage offered by coupled emitters is their capability to generate multi-photon states directly. Superradiant emitters have been proposed to generate states that offer increased phase resolution for quantum metrology~\cite{Paulisch2019}. Here, we implement on-demand two-photon state generation by fully exciting a pair of waveguide-coupled emitters, and study correlations and directionality in the emission~\cite{Maffei2024}.

\begin{figure*}
	\includegraphics[width=.8\linewidth]{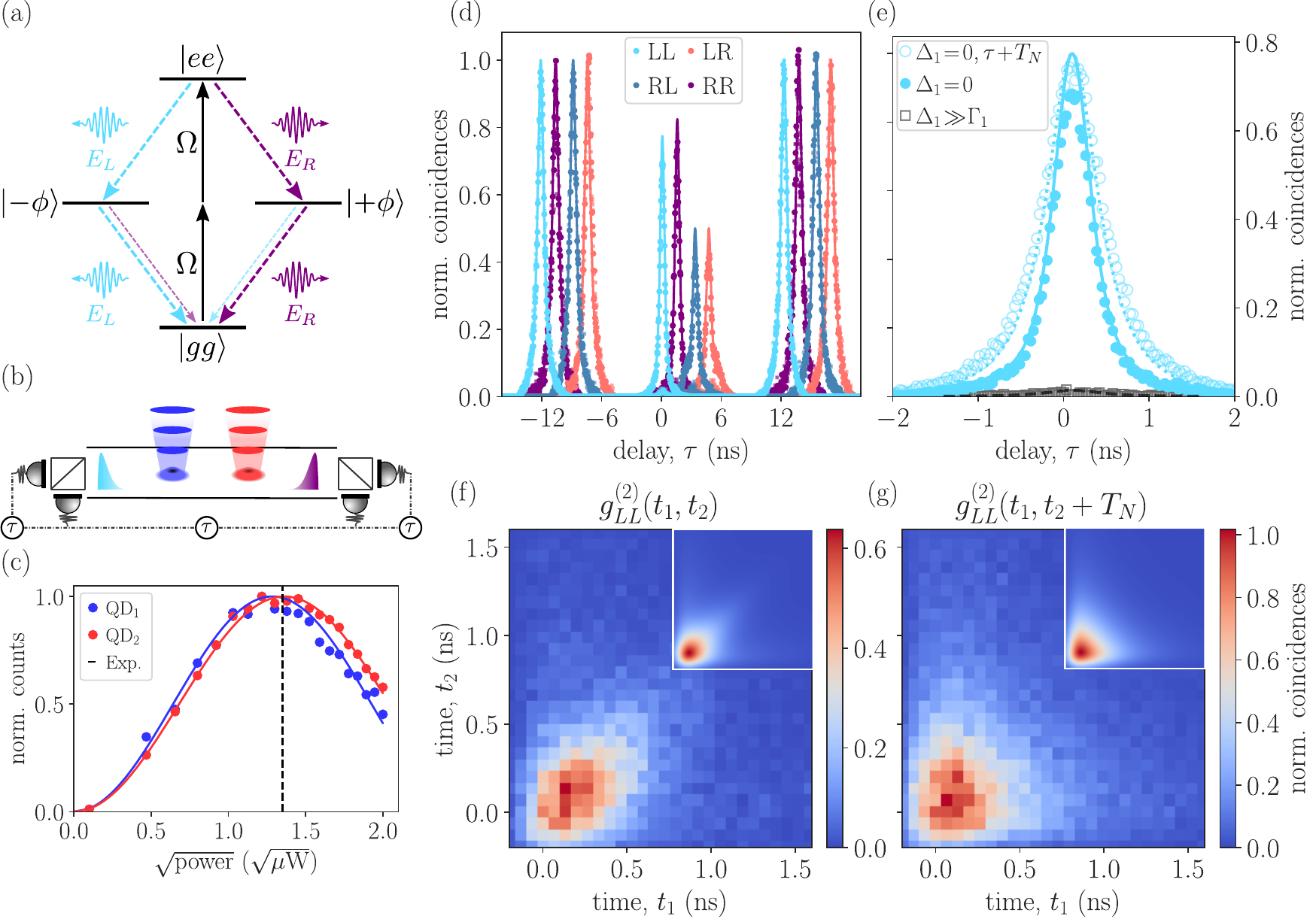}
	\caption{\textbf{Correlations in emission after full inversion.} 
    (a) Energy-level diagram for the coupled two-level emitters. Included is the double excited state, $\ket{ee}$ and the single-excitation subspace is expressed by the states $\ket{\pm\phi}=(\ket{eg} + e^{\pm i\phi}\ket{ge})/\sqrt{2}$. Excitation is indicated by solid arrows labeled with $\Omega$ and decay is shown by dashed arrows. Secondary arrows for decay to the ground state, $\ket{gg}$, are added to indicate the potential presence of a non-unidirectional emission, which depends on $\phi$. (b) Schematic of simultaneous resonant pulsed excitation of both QDs along with Hanbury Brown–Twiss setups for the output ports (see Appendix~\ref{app:muliplex-collection} for details). (c) Separately measured Rabi oscillations for QD$_1$ (blue) and QD$_2$ (red). Solid lines are fits with the Rabi model. The dashed line indicates the power used for the correlation measurements. (d) Measured intensity correlations, $g^{(2)}_{\alpha\beta}(\tau)$, for all four direction combinations when exciting both QDs (filled circles) and only QD$_2$ (open squares). The data for only exciting QD$_2$ is mainly visible as relatively low peaks near zero delay. $g^{(2)}_{LL}$ is centered at $\tau=0$ ns, while the other correlations are offset in the delay for clarity. Simulations of $g^{(2)}(\tau) = \int G^{(2)}(t,t+\tau) dt / \left(\int I(t) dt\right)^2$ are shown by solid lines. (e) Zoom in of (d) for the center peak of $g^{(2)}_{LL}$ (filled circles) shown with respect to the right y-axis. As reference a temporally far-away side peak is shown (open circles) with $\tau\to\tau+T_N$, where $T_N=500T$ with $T$ is the laser period. Again correlations for only exciting QD$_2$ are shown for comparison (gray open squares), which are close to the bottom of the figure. Lines show numerical simulations. (f), (g) Fully time-resolved correlations, $g^{(2)}_{LL}(t_1,t_2)$, of photons detected after (f) the same laser pulse and (g) different laser pulses. The coincidences are normalized to the maximum value in $g^{(2)}_{LL}(t_1,t_2+T_N)$. Insets show corresponding numerical simulations for the same axes as the experiment. 
    }
	\label{fig:4}
\end{figure*}

The coupled two-level emitters together form a four-level system, as shown by the energy level diagram in \fref{fig:4}(a). In contrast to \fref{fig:1}(e), we now explicitly include the double excited state, $\ket{ee}$, and the single excitation subspace is spanned by $\ket{\pm \phi}$. The full emission from $\ket{ee}$ can be considered as a two-step process. Specifically, in the case that the first photon is emitted to the left (right) port, the system decays to the $\ket{-\phi}$ ($\ket{+\phi}$) state. The direction of the subsequently emitted second photon is governed by the overlap of $\ket{+\phi}$ ($\ket{-\phi}$) with the state $\ket{\pi-\phi}$ ($\ket{\pi+\phi}$) from which emission is into only one direction as depicted in \fref{fig:1}(e). This results in temporal correlations between two photons emitted into either the same or opposite directions. 

In the experiment, schematically depicted in \fref{fig:4}(b), both QDs are simultaneously excited with a pulsed resonant laser using the SLM. For initialization of the coupled system into the $\ket{ee}$ state, i.e., to perform full inversion, first individual Rabi measurements are performed for each QD by detuning the other far from resonance. Figure~\ref{fig:4}(c) shows the Rabi oscillations as a function of laser power for each of the QDs measured separately. The overlap of the Rabi curves shows that the Rabi frequencies are approximately equal. Excitation into $\ket{ee}$ is then performed at the laser power corresponding to the $\pi$-flip for QD$_1$ and QD$_2$. 

Figure~\ref{fig:4}(d) shows the complete set of second-order intensity correlations, $g^{(2)}_{\alpha\beta}(\tau)$ with $\alpha, \beta \in \{L,R\}$, measured using HBT setups as shown in \fref{fig:4}(b). The center peaks consist of coincidences from the same laser pulse, while the side peaks correspond to coincidences from subsequent pulses. For clarity, the datasets for $g^{(2)}_{LR}$, $g^{(2)}_{RL}$, and $g^{(2)}_{RR}$ are displaced horizontally. For comparison, the correlations are shown for $\Delta_1\gg\Gamma_1$, effectively only addressing QD$_2$. This results in $g^{(2)}_{LL}(0)=0.01$, $g^{(2)}_{LR}(0)=0.05$, $g^{(2)}_{RL}(0)=0.02$, and $g^{(2)}_{RR}(0)=0.02$, obtained from the ratio between the maxima of the center to that of a side peak. This anti-bunching clearly demonstrates single-photon emission. With both QDs in resonance, the normalized center peak heights are $g^{(2)}_{LL}(0)=0.70$, $g^{(2)}_{LR}(0)=0.42$, $g^{(2)}_{RL}(0)=0.41$, and $g^{(2)}_{RR}(0)=0.76$. 

Similar to Eq.~\eqref{eqn:g2ll_pops} and Eq.~\eqref{eqn:g2rr_pops}, the correlations between the left and the right port are
\begin{equation}
g^{(2)}_{LR}(0) = g^{(2)}_{RL}(0) = \frac{p_{ee} \cos^2{\phi}}{\left(p_{ee} +p_{+\phi} \right)\left(p_{ee} +p_{-\phi} \right)}.
\end{equation}
From this set of equations, it follows that $g^{(2)}_{LR} = \sqrt{g^{(2)}_{LL}(0) g^{(2)}_{RR}(0)} \cos^2{\phi}$, which is consistent with the measured coincidences for a coupling phase $\phi=0.8 \pi$. Moreover, there is a clear contrast between photons going to the same port, $g^{(2)}_{LL}$ and $g^{(2)}_{RR}$, and to different ports, $g^{(2)}_{LR}$ and $g^{(2)}_{RL}$. This contrast in the multi-photon emission is referred to as emergent chirality, because it is a result of the spontaneously broken mirror symmetry~\cite{Cardenas-Lopez2023}. 

To further assess the temporal profile of the center peak, a zoom in of $g^{(2)}_{LL}(\tau)$ is shown in \fref{fig:4}(e). For comparison, a far-away side peak of $g^{(2)}_{LL}(\tau+T_N)$ is included when both QDs are excited, as well as the center peak when addressing only QD$_2$ ($\Delta_1\gg\Gamma_1$). As pointed out above, for the single excited QD the normalized coincidences are low. When exciting both QDs, the center peak is narrower compared to the side peak, which indicates that the emission is temporally correlated. 

The temporal correlations in the emission are captured in more detail by the fully time-resolved intensity correlation function, $g^{(2)}_{LL}(t_1,t_2)$. Figure~\ref{fig:4}(f) shows the correlation map with coincidences detected for the same laser pulse, while coincidences from different laser pulses are shown by the uncorrelated map in \fref{fig:4}(g). The diagonally elongated feature observed in the correlation map shows that two photons arrive together, while the corresponding feature is absent in the uncorrelated map. In essence, the probability distribution underlying the uncorrelated map is a product of the photon distributions for the independent emission after the two separate laser pulses. For the correlated map the distribution is not a product, because the distribution of the second photon depends on the detection time difference with the first photon. This is a consequence of the fact that the photon-pair emission is governed by a cascaded decay process. Alternatively, the emission of photon pairs, where two correlated photons travel together in the same direction, can be understood as one quantum emitter stimulating the emission from the other.

\section{Discussion}
This work demonstrates directional emission from a pair of independently tunable quantum dots coupled via a waveguide. In essence, the emitters form a waveguide-engineered molecule, which offers an alternative realization of chiral quantum optics~\cite{Gheeraert2020,Guimond2020}. The emission directionality can be increased by introducing on-chip phase shifters~\cite{Qvotrup2025}, and a secondary coupling mechanism that cancels the dispersive energy-level shifts while preserving the interference responsible for directional emission~\cite{Kannan2023}. Future work could involve exploiting the nonlinear phase shift induced by coupled emitters to perform conditional phase gates~\cite{Ralph2015}, and deterministic photon sorting~\cite{Yang2022}. Furthermore, the waveguide-engineered molecule could potentially serve as a crucial building block for quantum networks with directional light-matter interactions in waveguides~\cite{Mahmoodian2016}.

An exciting avenue opened up by the independent electric tuning is the operation of multiple QDs equipped with a spin degree of freedom~\cite{Warburton2013, Appel2022}. Interesting next steps are, for example, the deterministic generation of spin-spin entanglement~\cite{Gullans2012} and the implementation of quantum gates~\cite{Dzsotjan2010}, paving the way for the generation of two-dimensional cluster states~\cite{Economou2010,Osullivan2025}, a crucial resource for photonic quantum computing.

Zooming out, this work presents three key steps for the development of a multi-emitter waveguide QED platform based on QDs: (1) independent tuning of emission frequencies, which can be scaled by adding more trenches, (2) individual and collective optical control, which is readily scalable to a higher number of emitters by using tailored patterns for the spatial light modulator as demonstrated for optical atomic traps~\cite{Nogrette2014} and recently for multiple QDs in an unstructured medium~\cite{Shaji2026}, and (3) coupling over distances of several tens of wavelengths, thus extending the spatial range over which the emitters can be located. 

Two considerable challenges for scaling up are the random spatial locations of QDs and inhomogeneous broadening of an ensemble. To assess scalability, we have performed numerical simulations for an increased number of independently tunable waveguide regions, see Appendix~\ref{app:scal_feas}. The results demonstrate that sets of three and four resonant QDs are already feasible. Scaling with one additional emitter can be achieved by supplementing electric with magnetic tuning. The simulations further show that scaling to sets of ten can be enabled by a moderate few-fold increase in relative tunability and mean number of emitters per waveguide. The relative tunability can be increased by reducing the spectral distribution of the QD ensemble~\cite{Allen2000, Kersting2025} and by extending the tuning range with a heterostructure that has an additional doping layer~\cite{Lobl2017} or increased aluminum concentration~\cite{Bennett2010}. A higher mean number of emitters per structure can be realized with extended waveguides and growth at elevated QD density. In the future, scaling up may be expedited by developments in growth of QDs with pre-determined nucleation sites~\cite{McCabe2021, Han2021, Jons2013}. Nevertheless, even with randomly located emitters, deterministic fabrication by aligning the waveguide position with respect to spectrally and spatially pre-characterized emitters~\cite{Badolato2005, Pregnolato2020} will undoubtedly facilitate scaling. Note that the above discussion applies similarly to multiple independent QDs~\cite{Patel2010,Zhai2022,Papon2023}. All together, the results presented here constitute substantial progress for quantum dots and multi-emitter waveguide QED~\cite{Sheremet2023}, and establish a novel platform for many-body quantum optics~\cite{Carusotto2013,Chang2014,Noh2017}.

\section*{Acknowledgments}
\textbf{Funding:} The authors gratefully acknowledge financial support from Danmarks Grundforskningsfond (Grant No. DNRF139, Hy-Q Center for Hybrid Quantum Networks). O.A.D.S. acknowledges funding from the European Union’s Horizon 2020 research and innovation program under the Marie Skłodowska-Curie grant agreement no. 801199. N.B. and A.L. acknowledge financial support of the BMFTR QR. N project 16KIS2200, Eurostars BMFTR QTRAIN project 13N17328, QUANTERA BMFTR EQSOTIC project 16KIS2061, as well as DFG excellence cluster ML4Q project EXC 2004/1. \textbf{Competing interests:} P.L. is founder of the company Sparrow Quantum, which commercializes single-photon sources. The other authors declare no competing interests.

\bibliography{main.bib}

\clearpage
\onecolumngrid
\appendix

\section{Device, setup, and quantum dot locations}
\label{app:sample_setup}

\subsection{Device fabrication}
\begin{figure}[h!]
	\includegraphics[width=0.8\textwidth]{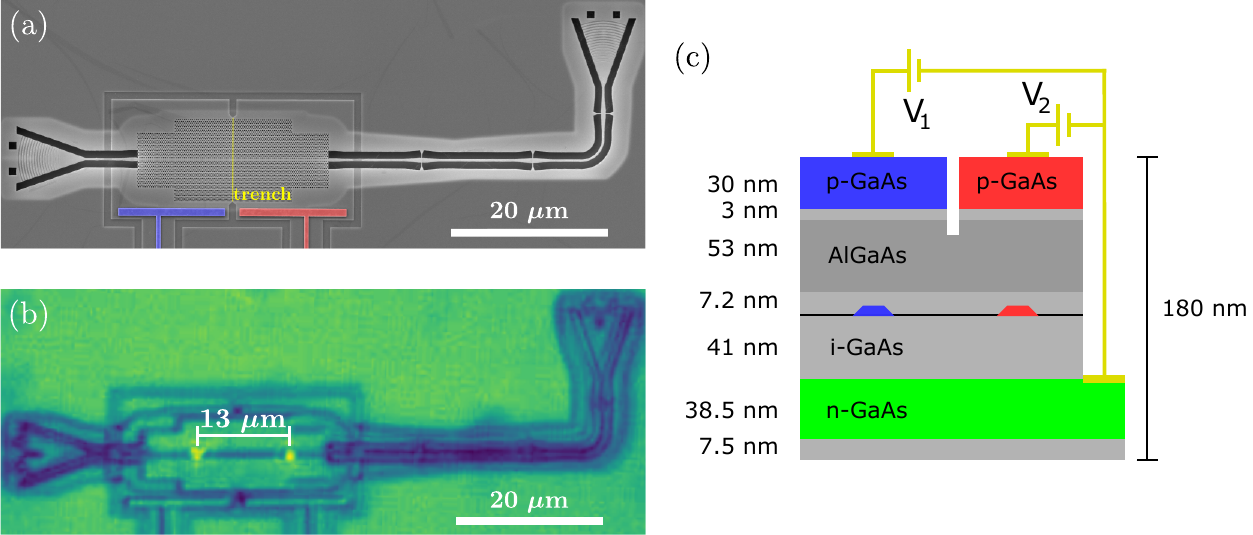}
	\caption{\textbf{Device and heterostructure.} (a) False-color coded scanning electron microscope image of a nominally identical device. Electrodes are color coded in blue and red and the shallow etched trench is labeled and colored in yellow. (b) Optical image of the device with QD positions indicated by the optimized beam spots. (c) Schematic of the membrane part of the heterostructure, which consists of a p-i-n diode and has the shallow etched trench through the p-doped layer. QDs are indicated by trapezoids.}
	\label{fig:dev_het}
\end{figure}

The device, see \fref{fig:dev_het}(a), consists of a photonic crystal waveguide with both ends connected to grating couplers via nanobeams. The two grating couplers couple out light with opposite polarizations facilitating separation in collection afterwards. A shallow trench is etched into the middle of the photonic crystal waveguide to enable independent electrical control for the left and right halves via separate electrical contacts. The trench is 80 nm wide and 60 nm deep, which was measured with atomic force microscopy. These dimensions are chosen such that the etch will be sufficiently deep for independent control while scattering due to the trench remains negligible. The shallow trench is etched together with the grating pattern of the couplers, and thus doesn't require an additional fabrication step. Electric tuning of the dipole transitions may enable increased coupling to the linearly polarized waveguide mode~\cite{Javadi2018}. The device is a suspended membrane that consists of a heterostructure based on GaAs with a p-i-n diode embedded, see \fref{fig:dev_het}(b). The shallow trench is etched only through the p-doped layer. The chip contains in total nearly 400 waveguide devices, with each of the electrical control lines shared by approximately 60 to reduce the required number of electrical wires and contacts. Further details about the fabrication process and growth can be found in Ref.~\cite{Uppu2020}. 

\subsection{Optical setup}
The device is placed on a printed-circuit board which is screwed onto a piezo stack and cooled down in a closed-cycle 4K cryostat with a confocal microscope configuration. On top of the cryostat is an optical breadboard on which the spatial light modulator is mounted. The breadboard furthermore holds linear polarizers, computer-controlled sets of quarter- and half-waveplates for the excitation and collection paths, powermeter heads to read the power for stabilization using a PID loop, camera and white light source for imaging of the device, and fiber couplers to couple light onto and out of the breadboard. In this experiment, the QDs are driven with either a continuously tunable continuous-wave laser or a pulsed fiber-optical parametric oscillator laser. The latter is filtered with a volume Bragg grating to suppress unwanted spectral components. The light coupled out via both ports is measured with superconducting-nanowire single-photon detectors (SNSPDs). 

\subsection{Emitter locations}
The positions of the QDs are extracted from a combination of three images: an image of the device, and two images each with an optimized beam spot location for one of the two QDs. The beam spots were independently optimized for excitation from free space on top of the waveguide in a resonant fluorescence scheme. A combination of the three images is shown in \fref{fig:dev_het}(b). 

\section{Characterization of the quantum dots}
\label{app:qd_char}

\subsection{Transmission spectrum}
\begin{figure}[h!]
	\includegraphics[width=.8\textwidth]{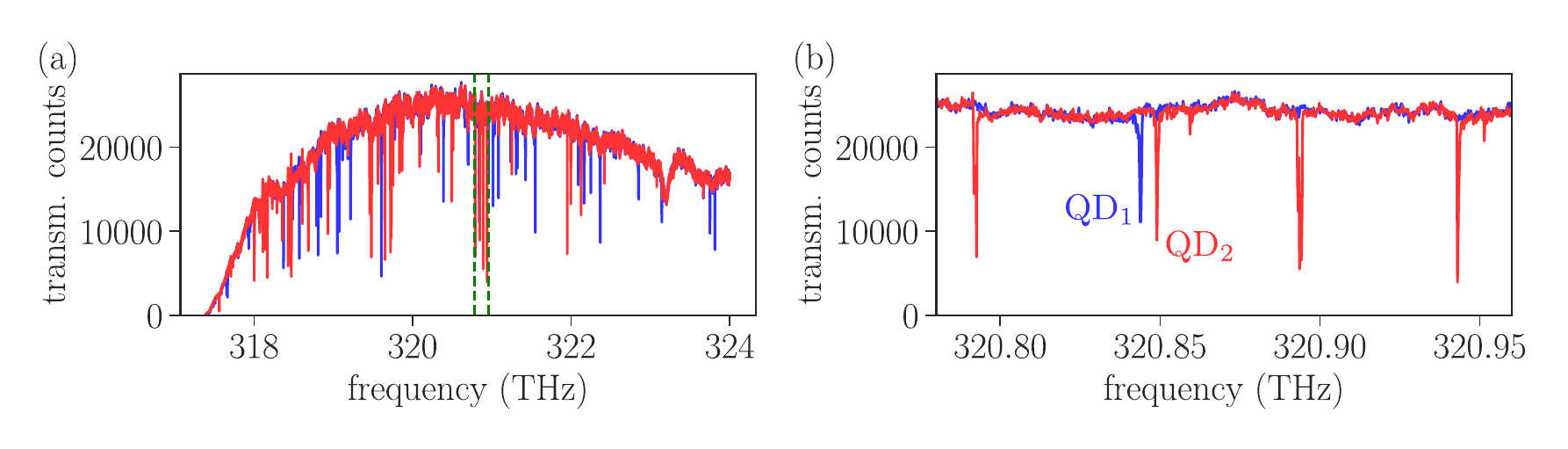}
	\caption{\textbf{Quantum dot search in transmission.} (a) Counts in transmission as a function of laser frequency with the left half bias voltage at $1.252$ V and the right half at $1.0$ V (blue), and vice versa (red). QDs appear as dips in the transmission spectrum. (b) Zoom in of the data in (a) for the window indicated by the dashed lines. Two dips with 5 GHz frequency difference are observed with one in each half of the waveguide. }
	\label{fig:bigrt}
\end{figure}

Figure~\ref{fig:bigrt}(a) shows two waveguide transmission spectra using a weak continuous wave laser. The spectra show that the photonic crystal band edge is around $317.4$ THz. The two spectra are taken with either one half or the other half of the waveguide at a voltage such that QDs can be resolved based on their radiative recombination. QDs in each half are identified via transmission dips. A pair of spectrally similar QDs is identified around $320.85$ THz in the zoom-in of the spectrum shown in \fref{fig:bigrt}(b). 

\subsection{Saturation in transmission}
\begin{figure}[h!]
    \centering
    \includegraphics[width=.8\linewidth]{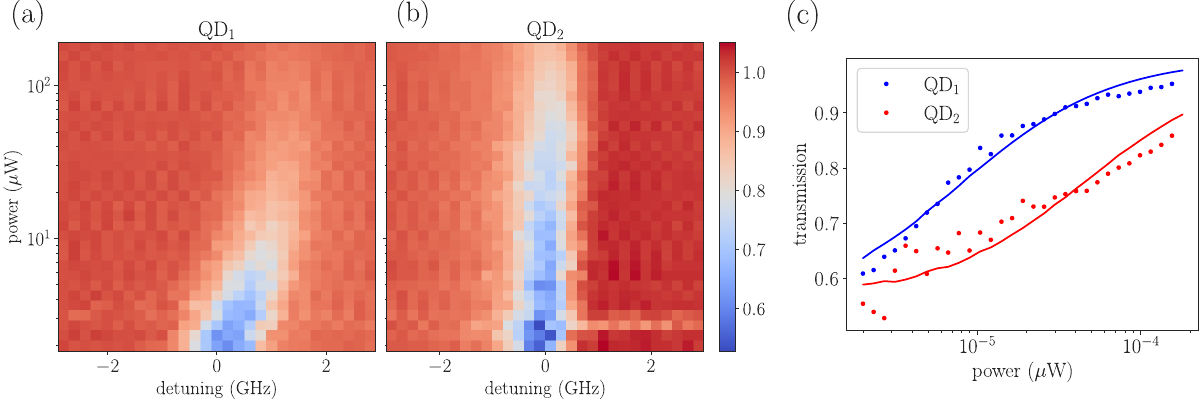}
    \caption{\textbf{Transmission and saturation.} (a), (b) Measured transmission intensity as a function of the input laser power and detuning of (a) QD$_1$ and (b) QD$_2$. (c) Minimum transmission dip for QD$_1$ (blue) and QD$_2$ (red). Solid lines are fitted with a model for the QD saturation~\cite{Thyrrestrup2018}. }
    \label{fig:RT}
\end{figure}

The QDs are further characterized by measuring the transmission intensity as a function of excitation power. Figure~\ref{fig:RT} shows the measured transmission as a function of both laser power and detuning for (a) QD$_1$ and (b) QD$_2$. The transmission values are normalized to the transmission obtained at large QD detuning. The measured minimum transmission dip as a function of excitation powers is shown in \fref{fig:RT}(c). We extract from these transmission measurements the pure dephasing, $\Gamma_d$, the Fano factors, $\xi_i$, which model the effect of a weak cavity caused by imperfect impedance matching~\cite{Thyrrestrup2018}, the couplings to the guided mode, $\beta_i$, and the spectral diffusions, $\sigma_{\text{sd}}$, and provide them in Table~\ref{tab:qd_pars}.

\subsection{Charge plateaus}
\begin{figure}[h!]
	\includegraphics[width=.67\textwidth]{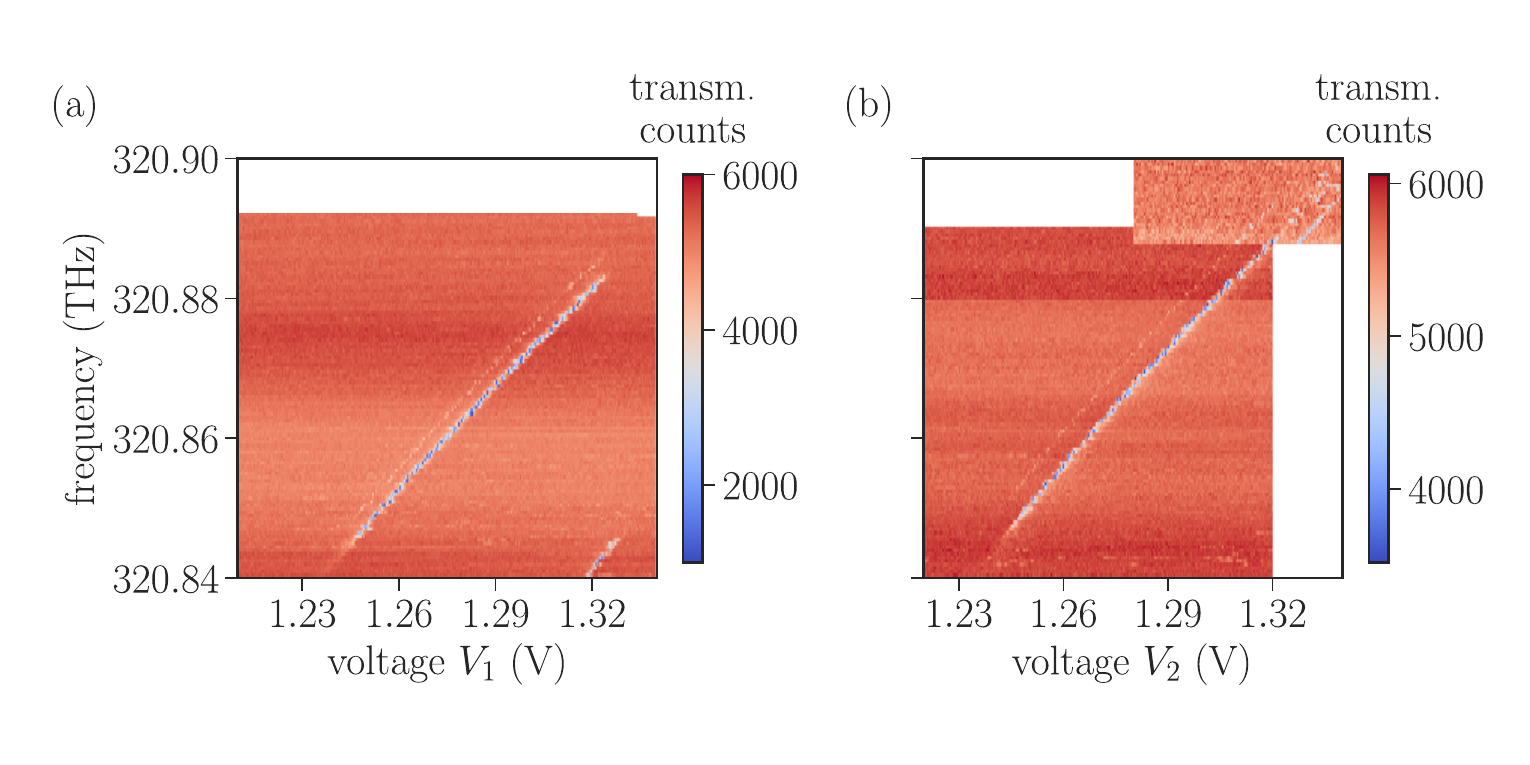}
	\caption{\textbf{Charge plateaus in transmission.} (a), (b) Counts as a function of laser frequency and bias voltage for the (a) left half, which holds QD$_1$, and the (b) right half, which holds QD$_2$, of the waveguide. The other bias voltage is set to $1.0$ V. The data in (a) is composed of three measurements, which had slightly different background signal. }
	\label{fig:chargeplat}
\end{figure}

Permanent dipoles and tuning ranges for each of the two QDs are obtained from the charge plateaus shown in \fref{fig:chargeplat}. Both QDs have one dipole that is well coupled to the waveguide, and for both the second dipole is only weakly coupled resulting in the faint additional lines. The electrical tuning range for both QDs is close to 50 GHz. Outside this range tunneling dominates over radiative recombination.

\subsection{Lifetime measurements}
\begin{figure}[h!]
    \centering
    \includegraphics[width=0.4\linewidth]{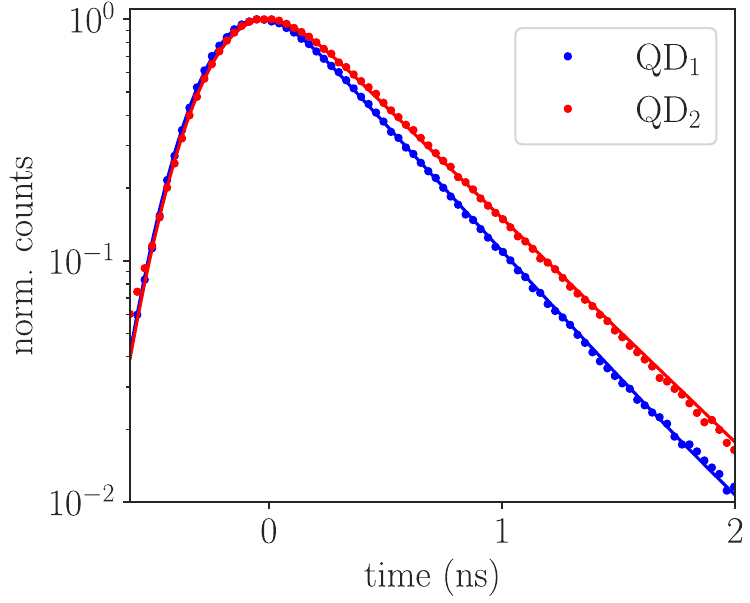}
    \caption{\textbf{QD lifetimes.} Normalized time-resolved emission intensity for QD$_1$ (blue) and QD$_2$ (red) for resonant pulsed excitation. Taking into account the instrument response function, the extracted exponential decay rates are $\Gamma_1/2\pi=0.388(1)$ GHz and $\Gamma_2/2\pi=0.349(1)$ GHz. }
    \label{fig:lifetimes}
\end{figure}

The lifetimes of the QDs are extracted experimentally by exciting them resonantly from the top with a pulsed laser. The emitted light is filtered by a $\approx 3$ GHz narrow etalon filter to remove leaked laser light, and is detected by an SNSPD. The measured time-resolved intensity signal, shown in \fref{fig:lifetimes}, is modeled by the convolution of the exponential QD decay with the instrument response function to take into the account the detection jitter. The instrument response function is a Gaussian distribution with standard deviation $\sigma_{IRF}=188(1)$ ps. The decay rates are $\Gamma_1/2\pi=0.388(1)$ GHz and $\Gamma_2/2\pi=0.349(1)$ GHz for QD$_1$ and QD$_2$. The parameters for the theoretical fits are presented in Table~\ref{tab:qd_pars}.

\begin{table}[h!]
\caption{\textbf{Parameters of the quantum dots.} QD total decay rates, $\Gamma_i$, the permanent dipole, $p_i$, for the electrical tuning by the dc Stark shift, $\xi_i$ the Fano factor, $\beta_i$ the coupling to the guided mode, $\sigma_{\text{sd}}$ the spectral diffusion, $\gamma_d$ the dephasing rates, and $\phi$ the coupling phase. }
\label{tab:qd_pars}
\setlength{\tabcolsep}{5pt}
\renewcommand{\arraystretch}{1.4}
\begin{tabular}{c|c}
\toprule[1pt]\midrule[0.3pt]
$\{\Gamma_1, \Gamma_2\}/2\pi$, GHz & 0.388(1), 0.349(1) \\
$\{p_1, p_2\}/2\pi$, GHz/mV & 0.50, 0.54  \\
$\{\beta_1, \beta_2\}$ & 0.95, 0.85 \\
$\{\xi_1, \xi_2\}$ & 0.0, 0.1 \\
$\sigma_{\text{sd}}/2\pi$, GHz  &0.30, 0.22    \\ 
$\gamma_d/2\pi$, GHz &0.01, 0.09   \\  
$\phi$, rad &  $0.8 \pi$ \\ 
\midrule[0.3pt]\bottomrule[1pt]  
\end{tabular}
\end{table}

\section{Spatial light modulator for excitation}
\label{app:slm}
We employ a spatial light modulator (SLM) to simultaneously excite both QDs from free space. The SLM is a Holoeye Pluto 2.1, which is for phase-only modulation. It consists of 1920 x 1080 pixels of 8 µm x 8 µm size and has a 93\% fill factor. Each pixel can be programmed individually to control the acquired phase with 8-bit resolution. 

\begin{figure}[h!]
    \centering
    \includegraphics[width=0.5\linewidth]{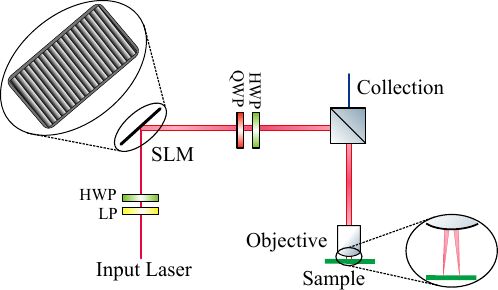}
    \caption{\textbf{Schematic of setup using SLM.} A linear polarizer (LP) and half-wave plate (HWP) align the input laser polarization for optimal SLM performance. A blazed grating is applied to the SLM to diffract the beam. A HWP and quarter-wave plate (QWP) tailor the polarization to excite a specific QD dipole. The beam passes through a beam splitter which separates excitation and collection, and the objective focuses it onto the device as two distinct spots. }
    \label{fig:slm_setup}
\end{figure}

For collective excitation, a blazed grating is programmed on the SLM to generate a diffraction pattern, see \fref{fig:slm_setup}. The zero order beam spot is positioned on QD$_1$ and the first order on QD$_2$ by tuning the period and rotation of the grating. The amplitude of the grating is used to control the relative driving strength, while the relative driving phase is controlled by translation of the grating.

\begin{figure}[h!]
	\includegraphics[width=\textwidth]{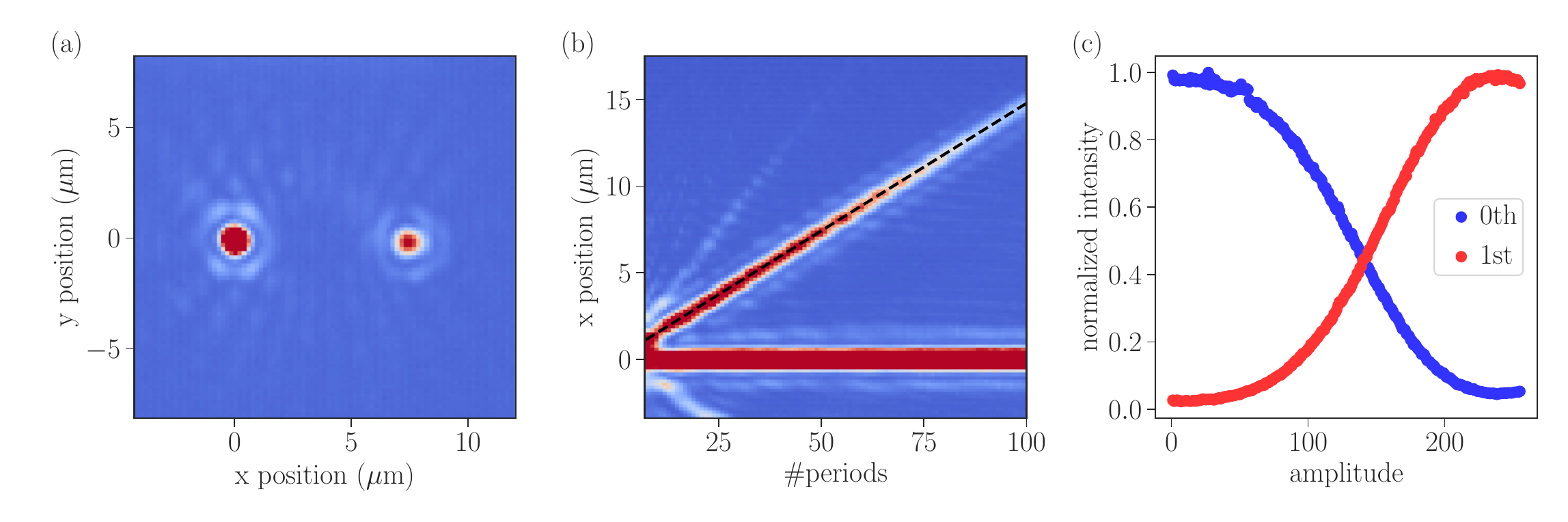}
	\caption{\textbf{Control of beam spots with SLM.} (a) Camera image with on the left the 0th and on the right the 1st diffraction order created with a blazed grating pattern on the SLM. (b) Slices at $y=0$ of camera images as a function of the number of periods patterned on the SLM screen. The dashed black line tracks the position of the 1st order diffraction. (c) Intensity of the 0th and 1st order diffraction beam spots as a function of the modulation depth of the blazed grating pattern. 
    }
	\label{fig:slm_control}
\end{figure}

Figure~\ref{fig:slm_control}(a) shows a representative picture taken with the camera used to image the device in the cryostat. The picture clearly shows two well-separated beam spots. Figure~\ref{fig:slm_control}(b) shows the change in position of the 1st order beam spot as a function of the number of grating periods programmed for the SLM. The position changes linearly with the number of periods, thus providing a convenient control knob for the alignment of the 1st order beam spot. The intensity of the 0th and 1st order beam spot as a function of the amplitude of the blazed grating are shown in \fref{fig:slm_control}(c). The intensity is initially fully in the 0th order, since there is no diffraction at low amplitudes. As the amplitude is increased, the intensity of the 0th order decreases, until nearly all intensity is diffracted into the 1st order. The maximal intensity for the 1st order beam spot is approximately equal to that of the 0th order, which shows that the diffraction occurs with high efficiency. 

In more detail, the procedure for collective excitation with the SLM is as follows. First, the 0th order spot is manually aligned to one of the QDs by adjustments to mirrors in the optical path. Next, starting with an intermediate amplitude for the blazed grating, the 1st order spot is aligned to the other QD by adjusting the number of periods and the rotation of the blazed grating programmed on the SLM screen. Then, the amplitude of the grating is calibrated to balance the intensity between the spots to, for example, apply equal Rabi driving strengths to both emitters. Finally, the relative driving phase is controlled with the translation of the grating to study the collective emission behavior of the QDs.

Future work, for example on several QDs equipped with spin, may leverage a single SLM operated with double modulation to independently control, in addition to the amplitude and phase, the polarization of multiple beam spots~\cite{Moreno2012}. 

\section{Setup to temporally multiplex collection}
\label{app:muliplex-collection}
\begin{figure}[h!]
	\includegraphics[width=0.6\textwidth]{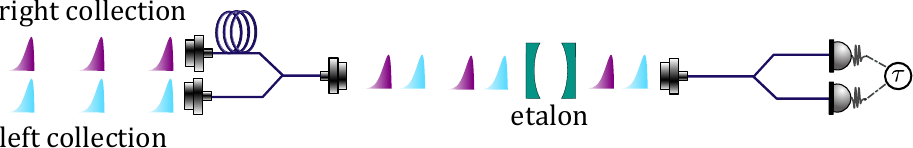}
	\caption{\textbf{Temporal multiplexing setup.} Schematic of setup to concurrently measure intensities, $I_L$ and $I_R$, and correlations, $g^{(2)}_{LL}$, $g^{(2)}_{RR}$, $g^{(2)}_{RL}$ and $g^{(2)}_{LR}$, of the left and right collection ports. The light from the right collection port is delayed and combined with the light from the left port on a beam splitter. The light is passed through an etalon filter and split up by another beam splitter after which the signals are detected by a pair of SNSPDs.}
	\label{fig:setup_col}
\end{figure}

To measure the emission intensities and their correlations under pulsed excitation, a setup is used to temporally multiplex the output signal as visualized in \fref{fig:setup_col}. One collection path is delayed with an additional fiber before combining the two paths with a 50:50 fiber beam splitter (FBS). The collected light is filtered together by a single narrow etalon and afterwards split up by a second 50:50 FBS. Thus, on each detector the filtered signal is recorded from both the left and right side of the waveguide, where the signal from the right output is delayed with respect to the left. In this way, for both outputs the intensities and all four combinations of intensity correlations can be measured with only one etalon filter and two SNSPDs.

\section{Directionality in emission intensity}
\label{app:dir_em}
The measurements underlying Fig.~\ref{fig:1}(f) are obtained by a lifetime-type measurement. Both QDs are resonantly excited using a single pulsed laser operated at relatively low power. In this way, emission is predominantly from the single-excitation subspace. The relative driving phase, $\theta_d$ is varied using the SLM (see Appendix~\ref{app:slm}). Figure~\ref{fig:dir_em} shows the recorded time-resolved emission intensities from the left and right output ports. The results demonstrate that the relative phase $\theta_d$ controls the emission directionality, directing photons preferentially to either the left or right output.
\begin{figure}[h]
	\includegraphics[width=0.6\textwidth]{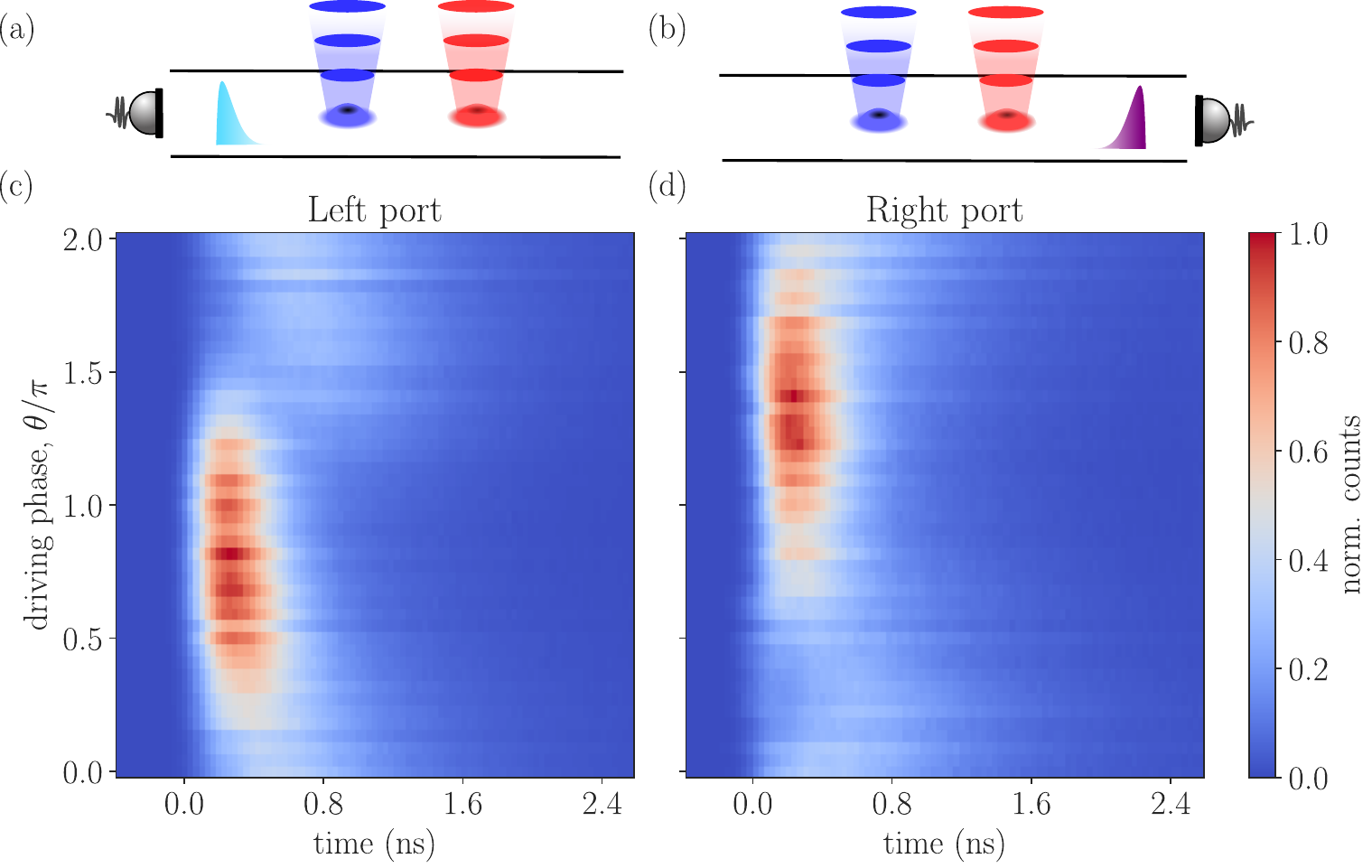}
	\caption{\textbf{Directionality in intensity as a function of driving phase} (a), (b) Schematics for pulsed excitation of both emitters and collection on the (a) left and (b) right port. (c), (d) Measured time-resolved emission intensity for the (c) left and (d) right port after pulsed resonant excitation. Counts are normalized by the maximum for the respective side. } 
	\label{fig:dir_em}
\end{figure}

\section{Further results for $g^{(2)}_{\alpha\beta}$ under continuous driving}
\label{app:g2_cw}

\subsection{Analytical derivations for $g^{(2)}_{\alpha\beta}(0)$}
In order to calculate the intensity correlations for a continuous drive at zero time delay $\tau =0$, we follow a weak-driving perturbation approach, which is recapped below and further described in Ref.~\cite{VanDiepen2025}. Furthermore, we extend the previous analysis by considering the directionality in photon statistics. 

Consider the state
\begin{align}
\ket{\psi} = \underbrace{c_{ee}}_{\sim \Omega^2} \ket{ee} + \underbrace{c_{eg} \ket{eg} + c_{ge}\ket{ge}}_{\sim \Omega^1} +  \underbrace{c_{gg}}_{\sim \Omega^0}\ket{gg}
\end{align}
where the coefficients will be expanded to the orders indicated above. From the effective Hamiltonian for the single-excitation subspace in the basis $\{\ket{eg},\ket{ge}\}$, 
\begin{align}
\mathcal{H}_\mathrm{eff} = \frac{1}{2}\left(
\begin{array}{cc}
 2\Delta_1 -i \Gamma _1 & -i e^{i \phi } \sqrt{\beta _1 \beta _2 \Gamma _1 \Gamma _2} \\
 -i e^{i \phi } \sqrt{\beta _1 \beta _2 \Gamma _1 \Gamma _2} & 2\Delta_2 -i \Gamma _2 \\
\end{array}
\right),
\end{align}
the steady state is calculated to leading order in $\Omega$. The solution for the steady state is 
\begin{align}
    \begin{pmatrix}
        c_{eg}\\ c_{ge}
    \end{pmatrix}_{ss} &= -\frac{\mathcal{H}_\text{eff}^{-1}}{2}\begin{pmatrix} \Omega_1 e^{i \theta_1 } \\ \Omega_2 e^{i \theta_2} \end{pmatrix}
\end{align}
and 
\begin{align}
    c_{ee,ss} &= \frac{i \left(e^{i \theta _2} \Omega _2 c_{\text{eg}}+e^{i \theta _1} \Omega _1 c_{\text{ge}}\right)}{\Gamma _1+\Gamma _2+2 i \left(\Delta _1+\Delta _2\right)}.
\end{align}

Let us consider the case that only one emitter is driven ($\Omega_1=\Omega$, $\Omega_2=0$), yet both are in resonance with the driving field ($\Delta_1=\Delta_2=0$). The emitters are further assumed to have identical decay rates ($\Gamma_1=\Gamma_2=\Gamma$) and equal coupling strength to the guided mode ($\beta_1=\beta_2=\beta$). In this limit, the single-excitation subspace component of the steady state is
\begin{equation}
\ket{\psi_1} = \frac{i \Omega}{\Gamma B_\phi} \left( \ket{eg} - \beta e^{i\phi} \ket{ge} \right),
\end{equation}
with $B_\phi = 1 - \beta^2 e^{2 i \phi}$. As described in the main text, for $\beta \to 1$, $\ket{\psi_1} \propto \ket{\pi + \phi}$, thus single photons are emitted predominantly to the left. 

Experimentally, the directionality in photon statistics is characterized through the normalized intensity correlations, because those are not affected by differences in collection efficiencies. 
The intensity correlations in normalized form are $g^{(2)}_{\alpha\beta}(\tau) = G^{(2)}_{\alpha\beta}(\tau) / (I_\alpha I_\beta)$ with $G^{(2)}_{\alpha\beta}$ the unnormalized correlations and $I_\alpha$ the intensity for directions $\alpha,\beta \in \{L,R\}$. Explicitly, the intensities for the left and right port, $I_\alpha=\langle E^\dagger_\alpha E_\alpha \rangle$, are
\begin{align}
I_L & = \frac{\beta \Omega^2 \left( 2 (1+\beta)^2\Gamma^2 + \beta^2 \Omega^2 - 4\beta \Gamma^2 \cos 2 \phi \right)}{4\Gamma^3 |B_\phi|^2} , \\
I_R & = \frac{\beta \Omega^2 \left( 2 (1-\beta)^2\Gamma^2 + \beta^2 \Omega^2 \right)}{4\Gamma^3 |B_\phi|^2}.
\end{align}

The unnormalized intensity correlations, $G^{(2)}_{\alpha\beta}(0)=\langle E_\alpha^\dagger E_\beta^\dagger E_\beta E_\alpha \rangle$, are
\begin{align}
G^{(2)}_{LL}(0) & = G^{(2)}_{RR}(0) = \frac{\beta^2 \Omega^4}{4 \Gamma^2 |B_\phi|^2}, \\
G^{(2)}_{LR}(0) & = G^{(2)}_{RL}(0) = \frac{\beta^2 \Omega^4 \cos^2 \phi}{4 \Gamma^2 |B_\phi|^2} .
\end{align}
For $\phi=0$ the correlations are equal for the left and the right port. By increasing $\phi$, the correlations decrease for the left port and are minimal for $\phi=\frac{\pi}{2}$, while the correlations at the right port increase and are maximal at $\phi=\frac{\pi}{2}$. 

Regarding the photon statistics, in this work experimentally characterized by $g^{(2)}_{\alpha\alpha}(\tau)$, both the single-photon and the two-photon contribution change as a function of $\phi$. However, the directionality in $g^{(2)}_{LL}(0)$ and $g^{(2)}_{RR}(0)$ is caused by the change in single-photon component with the coupling phase. The directionality in the two-photon component could become apparent in the temporal asymmetry of $g^{(2)}_{LR}$. 
Our analytical result, however, only applies to $\tau=0$ and we therefore resort to a numerical investigation. 
\begin{figure}[h!]
	\includegraphics[width=.85\linewidth]{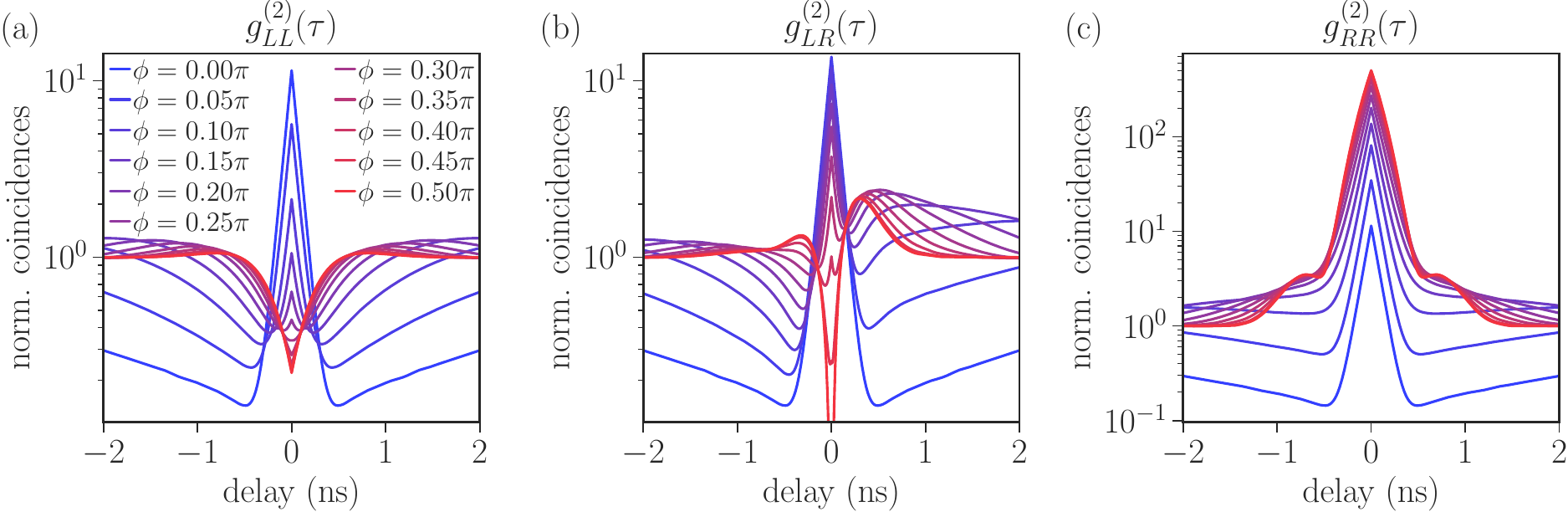}
	\caption{\textbf{Directionality in statistics.} 
    Numerically simulated intensity correlations (a) at the left port, (b) between left and right ports, and (c) at the right port for different coupling phases. The left emitter is excited with a weak resonant continuous drive and the right emitter is in resonance but not driven. Specifically, the simulation parameters are $\Omega_1/2\pi= 1/16$ GHz, $\Omega_2=0$ GHz, $\beta_1=\beta_2=0.95$, $\Gamma_1/2\pi=\Gamma_2/2 \pi=1$ GHz, and with dephasing for both emitters $\gamma_d/2\pi = .01$ GHz. These simulations do not include detection jitter and spectral diffusion.
    }
	\label{sfig:cw_g2_sim}
\end{figure}

\subsection{Numerical simulations of $g^{(2)}_{\alpha\beta}(\tau)$}
Simulations of the intensity correlations as a function of delay $\tau$ for various coupling phases and different detector combinations are shown in \fref{sfig:cw_g2_sim}. For purely dissipative coupling, $\phi = N \pi$, there is no directionality in the intensity correlations. 
The directionality increases with increasing dispersive coupling. For purely dispersive coupling, the simulated intensity correlations show clear antibunching on the side of the driven emitter, while maximal bunching appears on the side of the not driven emitter. In addition, antibunching occurs in the correlations between the left and right ports. This is explained by that detection of the first photon projects the system into a state from which the second photon is emitted in the same direction. The oscillations are due to dynamics in the single-excitation subspace induced by the dispersive coupling. 

\subsection{Collective driving measurements}
\begin{figure}[h!]
    \includegraphics[width=.6\linewidth]{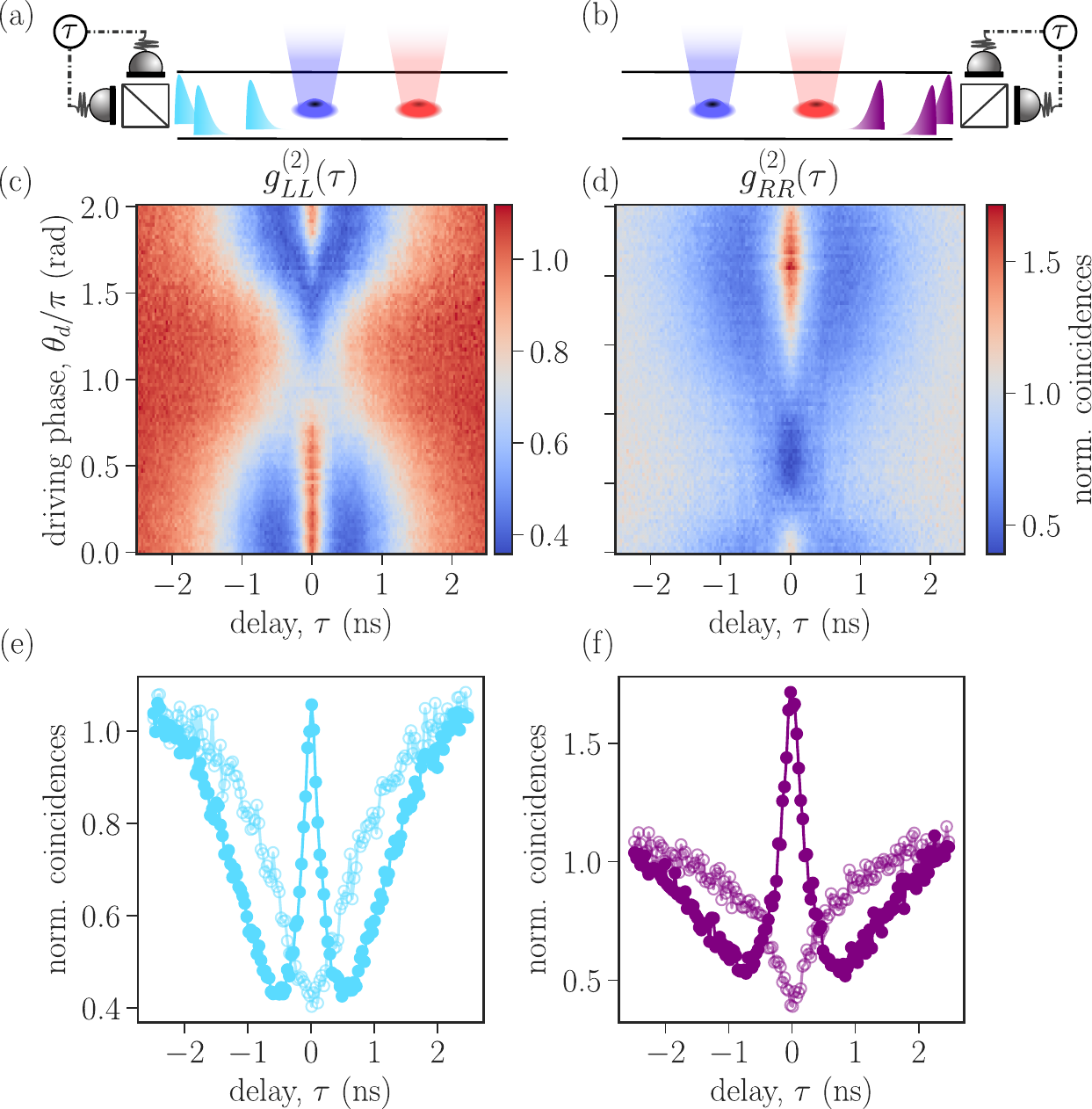}
	\caption{ \textbf{Directionality in photon statistics for collective driving.} 
    (a), (b) Schematic of a pair of continuously driven emitters in a waveguide with either the (a) left or the (b) right port leading to a Hanbury Brown-Twiss setup, which equally splits the scattered light into two detectors. (c), (d) Intensity correlations measured at the (c) left and (d) right port as a function of the relative driving phase. (e), (f) Cuts from (c) and (d), at the driving phase for the maximal (filled circles) and minimal (open circles) values of $g^{(2)}_{\alpha\alpha}(0)$. 
    }
	\label{fig:cw_g2_both}
\end{figure}

Figure~\ref{fig:cw_g2_both} shows coincidence measurements for weak and continuous resonant driving of both emitters. For the left port, see \fref{fig:cw_g2_both}(c) and (e), the maximum is $g^{(2)}_{LL}(0)=1.05$ while the minimum is $g^{(2)}_{LL}(0)=0.40$. At the right port, see \fref{fig:cw_g2_both}(d) and (f), the maximum is $g^{(2)}_{RR}(0)=1.71$ while the minimum is $g^{(2)}_{RR}(0)=0.39$. Thus on both the left and the right port, the emission changes between antibunching and bunching as the relative driving phase between the emitters is varied. Notably, there are driving phases where either the left port, the right port, or both ports show an antidip in the coincidences. These results demonstrate the role of coherence in the driving of both emitters. 

\section{Scalability analysis}
\label{app:scal_feas}

\subsection{Success conditions and assumptions}
In this section we assess the feasibility of scaling up the system of coupled QDs based on individual electrical control enabled by employing multiple trenches. More specifically, we calculate probabilities of finding at least one set of radiatively coupled QDs that can all be tuned into resonance simultaneously. To this end, three conditions must be met: each QD in the set must be (1) well coupled to the same waveguide, (2) individually tunable, thus separated by trenches from the other QDs in the set, and (3) spectrally sufficiently close to the other QDs in the set, such that they can all be tuned into resonance simultaneously.

To calculate the probability of satisfying these conditions, a statistical model is implemented with several assumptions. For condition (1), a device is considered with a certain QD density, $\rho_{QD}$, and with the locations of the QDs independent and identically distributed. For waveguides, fabricated without taking into account the emitter locations, the number of QDs in it follows a Poisson distribution with mean, $\mu_{QD}=\rho_{QD} A$, where $A$ is the area of the waveguide. A subset of these QDs will be well coupled to the waveguide, depending on its geometry and the locations of the QDs within it. To obtain a reference, we count the number of well-coupled QDs in the structure used in this work, conditioned on their resonant transmission dip being deeper than 50\%. This results in 35 QDs, see \fref{fig:bigrt}(a), thus $\mu_{QD}=35$ is taken as a representative value for the current device. For (2), it follows that the subset of QDs in each of the electrically independent regions in the waveguide, $N_{reg}$, is given by the Poisson distribution. Regarding (3), the inhomogeneous broadening of the emission wavelengths is modeled by a Gaussian with width $\sigma_{QD}=15$ nm. From the charge plateaus shown in \fref{fig:chargeplat}, the tuning range of the emission wavelength, $\delta\lambda$, is extracted to be approximately $0.15$~nm, which determines the spectral separation condition.

\subsection{Probability by sampling}
\begin{figure}[ht]
    \centering
    \includegraphics[width=.7\linewidth]{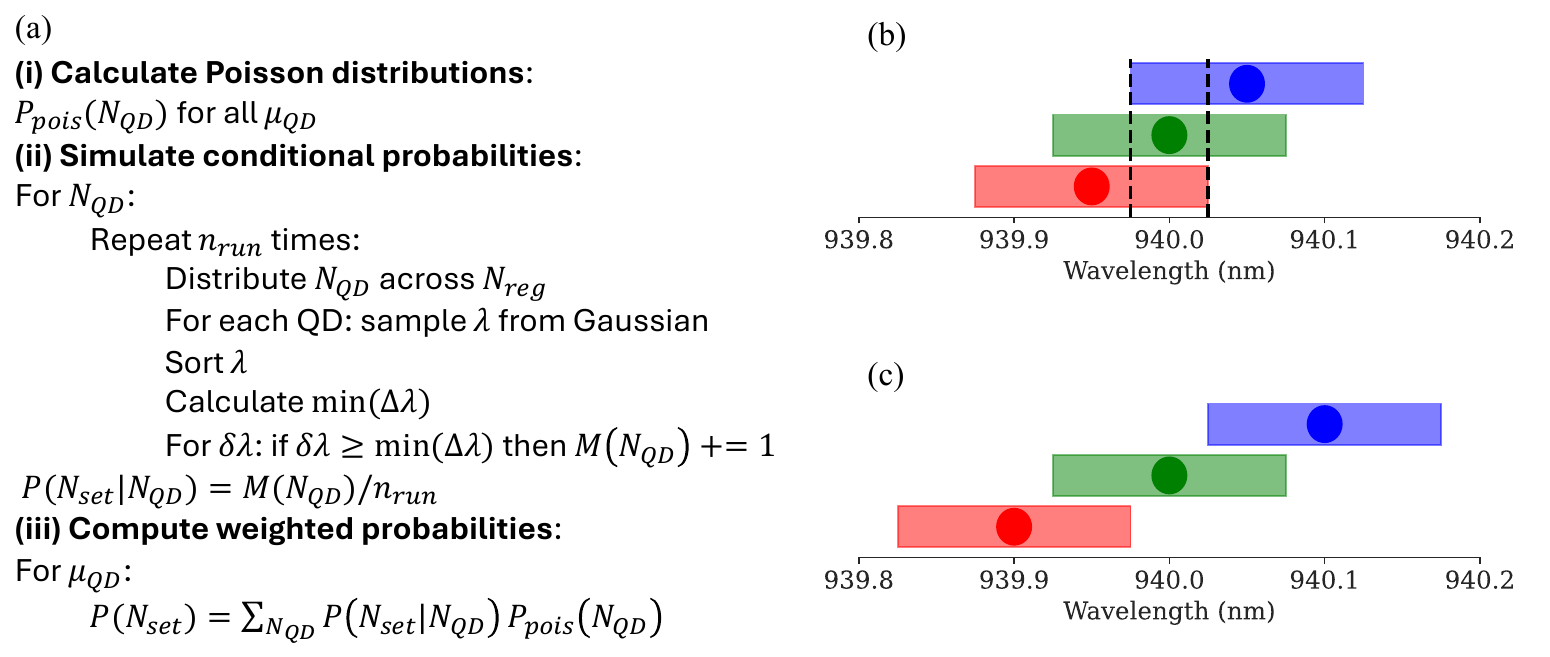}
    \caption{\textbf{Algorithm and wavelength condition.} (a) Pseudocode of the Monte Carlo sampling algorithm for a certain $N_{set}$. (b) 3 independently tunable QDs that are spectrally sufficiently close such that the full set can be tuned into resonance simultaneously. The dashed lines indicate the range of wavelengths for which they can be made resonant. (c) 3 QDs being spectrally too dissimilar to all be within tuning range of each other.} 
    \label{fig:MonteCarloAlgorithm}
\end{figure}

Given these three conditions, the goal is to extract the probability of finding a set containing the number of QDs, $N_{set}$. This probability is obtained from an algorithm based on Monte Carlo sampling, shown as pseudocode in \fref{fig:MonteCarloAlgorithm}(a). 

In the first step (i), we calculate the Poisson probability amplitude, $P_{pois}(N_{QD})$, to find a number of QDs, $N_{QD}$, in a waveguide for a given mean, $\mu_{QD}$. The Poisson distribution is calculated up to a $N_{QD}$ that includes at least 99.95\% of the probability mass. In the second step (ii), the conditional probability, $P(N_{set}|N_{QD})$, is computed for every $N_{QD}$ by repeatedly running the sampling routine. For each run, first, the QDs are distributed across $N_{reg}$ regions following the multinomial distribution. Next, for each QD its emission wavelength is sampled from a Gaussian distribution. Then, the wavelengths are sorted and the smallest spread, $\min(\Delta\lambda)$, is extracted for $N_{set}$ consecutive wavelengths that are for QDs in distinct waveguide regions. If $\delta\lambda \ge \min(\Delta\lambda)$, then a success is noted. This condition is visualized for $N_{set}=3$ in \fref{fig:MonteCarloAlgorithm}(b) and (c) for a success and a failure, respectively. Last, the number of successes is normalized to the total number of runs to obtain the conditional probability, $P(N_{set}|N_{QD})$. The third step (iii) is to combine the sampled conditional probabilities with the Poisson probability amplitudes, resulting in the final probability
\begin{equation}
    P(N_{set}) = \sum_{N_{QD}} P(N_{set}|N_{QD})P_{pois}(N_{QD}),
\end{equation}
for every given $\delta\lambda$ and $\mu_{QD}$.

A single chip typically holds many waveguides, which leads to the probability of finding at least one set of $N_{set}$ QDs that can be tuned into resonance on a chip being
\begin{equation}
    P_{N_{wg}}(N_{set}) = 1 - (1-P(N_{set}))^{N_{wg}},
\end{equation}
with $N_{wg}$ the number of waveguides.

\subsection{Scalability results}
\begin{figure}[h!]
    \centering
    \includegraphics[width=\linewidth]{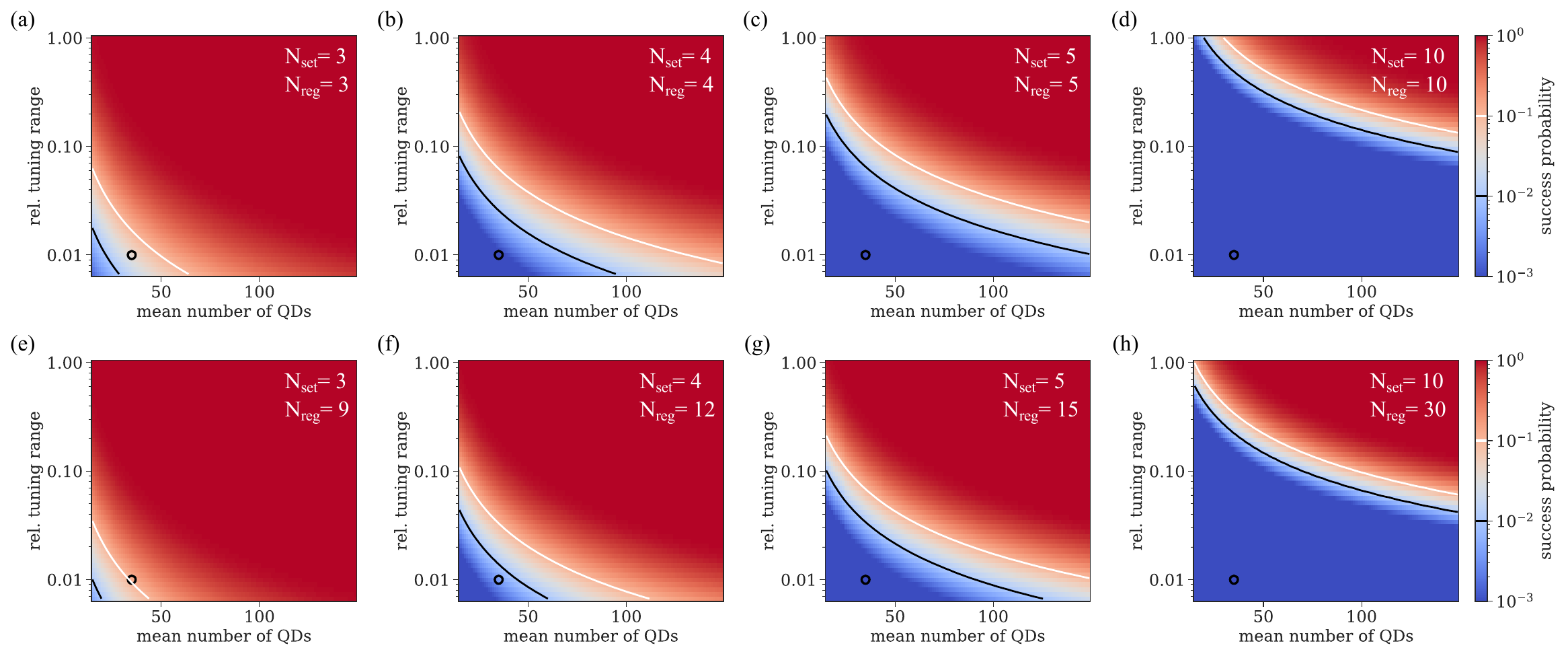}
    \caption{\textbf{Simulated probabilities per single waveguide.} (a)-(d) Probability of finding $N_{set}$ QDs in a single waveguide that can be tuned into resonance, for given relative tuning range, $\delta\lambda/\sigma_{QD}$, and mean number of QDs, $\mu_{QD}$. In the top right of each panel we indicate the size of the set and the number of waveguide regions. Black circles indicate the parameters of the device used in this work. The contour lines indicate the probability of 1\% (black) and 10\% (white). (e)-(h) Same as (a)-(d) with an additional number of trenches such that $N_{reg}=3N_{set}$. }
    \label{fig:Scalability_heatmaps}
\end{figure}

Figure~\ref{fig:Scalability_heatmaps} shows the simulated probabilities for a single waveguide, $P_1(N_{set})$, for several values of $N_{set}$ as a function of the relative tuning range, $\delta\lambda/\sigma_{QD}$, and the mean number of QDs, $\mu_{QD}$. In \fref{fig:Scalability_heatmaps}(a), for the current experimental parameters, a value of $P_1(N_{set}=3)=0.04$ is found. For $N_{set}=3$ this results in a probability of $P_{100}(N_{set}=3) = 0.98$ for a chip with 100 waveguides, thus it already becomes feasible to find sets of three resonant QDs with waveguides that have two trenches. For scaling further to $N_{set}=4$, shown in \fref{fig:Scalability_heatmaps}(b), with the current experimental settings, the probability for a single waveguide is $P_1(N_{set}=4)= 7\cdot 10^{-4}$ and for a chip with $N_{wg}=500$ it is $P_{500}(N_{set}=4) = 0.29$. This shows that it is in principle possible to operate sets of four resonant QDs with waveguides that have three trenches, but with this approach it likely requires to search multiple chips. Scaling to even larger sets of QDs is shown in \fref{fig:Scalability_heatmaps}(c) and (d), which for current experimental parameters shows $P_{500}(N_{set}=5)= 5 \cdot 10^{-3}$ and still lower for $N_{set}=10$.

The probability for larger sets of resonant QDs can be increased by separating a waveguide into more independently tunable regions by adding trenches. Intuitively, additional trenches enable the tuning into resonance of QDs that would otherwise not be independently tunable. Figure~\ref{fig:Scalability_heatmaps}(e)-(h) show the probabilities for three times the number of independently tunable regions, i.e., $N_{reg}=3N_{set}$. For current experimental settings, the probability for $N_{set}=4$, see \fref{fig:Scalability_heatmaps}(f), is increased to $P_1(N_{set}=4)= 4 \cdot 10^{-3}$, which corresponds to a factor three increase in probability per chip to $P_{500}(N_{set}=4) = 0.87$. For $N_{set}=5$, the increase in number of regions results in that the probability per chip is ten times that before so that $P_{500}(N_{set}=5)= 0.05$. However, this probability remains rather low thus such set sizes are challenging to realize given the current experimental parameters. 

Larger sets of QDs become feasible with increased relative tuning range and mean number of QDs, especially when combined with an increased number of trenches. Here we set the number of regions to be three times the set size, as shown in \fref{fig:Scalability_heatmaps}(e)-(h). Only twice the relative tuning range and mean number of QDs, $P_1(N_{set}=5)= 4 \cdot 10^{-3}$, results in $P_{500}(N_{set}=5)= 0.86$ per chip. With five times the relative tuning range and three times the mean number of QDs, we find $P_1(N_{set}=5)= .95$, i.e., a high probability already for a single waveguide. For even larger sets, say $N_{set}=10$, this results in the probability per waveguide $P_{500}(N_{set}=10)= 1 \cdot 10^{-3}$, which per chip is $P_{500}(N_{set}=10)= 0.42$. 

\begin{figure}[ht]
    \centering
    \includegraphics[width=\linewidth]{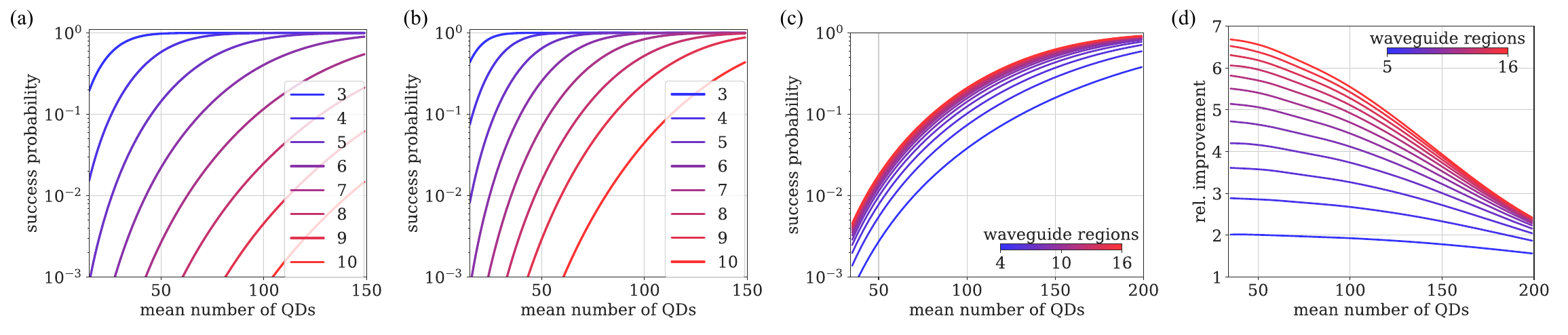}
    \caption{\textbf{Scaling with extra trenches.}  (a), (b) The probability that $N_{set}$ QDs, labeled in the legend, in a waveguide can be tuned into resonance as a function of the mean number of QDs. The relative tuning range is $\delta\lambda/\sigma_{QD} = 0.1$. In (a) the number of trenches is the minimally required, i.e. $N_{reg} = N_{set}$, and in (b) an excess number of trenches is used such that $N_{reg} = 3N_{set}$. 
    (c) Probability and (d) relative probability improvement for $N_{set}=4$ and various numbers of trenches, using the relative tuning range of $\delta\lambda/\sigma_{QD}=0.01$. }
    \label{fig:Scalability_trenches_and_setsize}
\end{figure}

Figure~\ref{fig:Scalability_trenches_and_setsize} presents the trend of the probability for various set sizes and numbers of trenches. For the minimum number of regions, i.e. $N_{reg} = N_{set}$, \fref{fig:Scalability_trenches_and_setsize}(a) clearly indicates that an increased mean number of QDs enables larger sets of resonant QDs. In comparison, a substantial increase in probability is obtained for additional trenches, as shown for $N_{reg} = 3N_{set}$ in \fref{fig:Scalability_trenches_and_setsize}(b). Specifically, for scaling to $N_{set}=4$, \fref{fig:Scalability_trenches_and_setsize}(c) shows the probabilities for various numbers of waveguide regions. Figure~\ref{fig:Scalability_trenches_and_setsize}(d) displays the relative improvement obtained from the data in (c) by normalization to the probability for $N_{reg} = N_{set}$. The improvement from separating into more independent regions decreases for an increasing number of regions, which can be explained by the fact that many QDs are already independently tunable. From \fref{fig:Scalability_trenches_and_setsize}(c) and \fref{fig:Scalability_trenches_and_setsize}(d) together, it is clear that the largest improvement obtained by adding trenches is in the low-probability regime. 

\subsection{Conclusion and outlook}
To sum up, the simulations demonstrate that waveguides with trenches for independent tuning, as presented in this work, are suited for scaling the number of coupled QDs. The first step in scaling is three resonant QDs. Our feasibility analysis shows that this is well within reach based on the probability of $0.98$ for a chip with 100 waveguides that have three independently tunable regions, i.e., with two trenches. An alternative approach may leverage a global magnetic field as an additional degree of freedom for tuning, therefore even the current device design with only one trench already enables three resonant QDs.

Furthermore, the feasibility analysis shows that additional trenches can enable scaling. Leveraging this approach, scaling to four QDs is realistic using twelve independent regions based on the simulated probability of $0.87$ for a chip with 500 waveguides. Further scaling to larger sets, e.g., ten waveguide-coupled QDs and beyond, becomes feasible with realistic adjustments to inhomogeneous broadening, tuning range, QD density, and waveguide dimensions. A powerful approach for scaling not included in our feasibility analysis, is deterministic fabrication through which the position of the waveguides is optimized with respect to pre-located emitters~\cite{Badolato2005,Pregnolato2020}. In the future, developments in growth of QDs with pre-determined nucleation sites~\cite{McCabe2021,Han2021,Jons2013} may enable considerable further scaling.

\end{document}